\documentclass[prd,nofootinbib,superscriptaddress]{revtex4}

\usepackage[british]{babel}
\usepackage{amsmath}
\usepackage{amssymb}
\usepackage{amsfonts}
\usepackage{graphicx}

\renewcommand{\eqref}[1]{Eq.~(\ref{#1})}

\newcommand{\be}{\begin{equation}}
\newcommand{\ee}{\end{equation}}
\newcommand{\ben}{\begin{equation*}}
\newcommand{\een}{\end{equation*}}
\newcommand{\bea}{\begin{eqnarray}}
\newcommand{\eea}{\end{eqnarray}}
\newcommand{\bean}{\begin{eqnarray*}}
\newcommand{\eean}{\end{eqnarray*}}
\newcommand{\brr}{\begin{array}}
\newcommand{\err}{\end{array}}
\newcommand{\bc}{\begin{center}}
\newcommand{\ec}{\end{center}}

\newcommand{\cf}{\mbox{\it cf.~}}
\newcommand{\vev}[1]{\mbox{$\langle #1 \rangle $}}
\newcommand{\leqsim}{\,\raisebox{-0.6ex}{$\buildrel < \over \sim$}\,}

\newcommand{\bk}{{\mathbf k}}

\newcommand{\bp}{{\mathbf p}}

\newcommand{\bx}{{\mathbf x}}

\newcommand{\bq}{{\mathbf q}}

\newcommand{\II}{\mathcal{I}}

\newcommand{\al}{\alpha}

\newcommand{\de}{\delta}

\newcommand{\la}{\lambda}
\newcommand{\Om}{\Omega}

\newcommand{\Gauss}{{\mbox{Gauss}}}

\newcommand{\tk}{\tilde{k}}  
 \newcommand{\tK}{\tilde{K}}
 \newcommand{\tbk}{\tilde{\bk}} 
\def\be{\begin{equation}}
\def\ee{\end{equation}}
\def\ba{\begin{eqnarray}}
\def\ea{\end{eqnarray}}

\begin{document}


\title{The Cosmic Microwave Background Temperature Bispectrum from Scalar Perturbations Induced by Primordial 
Magnetic Fields}

\author{Chiara Caprini}
\email{chiara.caprini@cea.fr} \affiliation{CEA, IPhT \& CNRS, URA 2306, F-91191 Gif-sur-Yvette, France}

\author{Fabio Finelli} 
\email{finelli@iasfbo.inaf.it} \affiliation{INAF-IASF Bologna, via Gobetti 101, I-40129 Bologna, Italy}
\affiliation{INAF-OAB, Osservatorio Astronomico di Bologna, via Ranzani 1, I-40127 Bologna, Italy}
\affiliation{INFN, Sezione di Bologna,
Via Irnerio 46, I-40126 Bologna, Italy}

\author{Daniela Paoletti}
\email{paoletti@iasfbo.inaf.it} \affiliation{Dip. di Fisica, Universit\`a degli studi di Ferrara and INFN,
via Saragat 1, I-44100 Ferrara, Italy}
\affiliation{INAF-IASF Bologna, via Gobetti 101, I-40129 Bologna, Italy}

\author{Antonio Riotto}
\email{antonio.riotto@cern.ch} \affiliation{CERN, PH-TH Division, CH-1211 Geneva 23, Switzerland}
\affiliation{INFN, Sezione di Padova, via Marzolo 8, Padova I-35131, Italy}

\date{\today}

\begin{abstract}
We evaluate the angular bispectrum of the CMB temperature anisotropy at large angular scale 
due to a stochastic background of primordial magnetic fields. The shape of non-Gaussianity depends
on the spectral index of the magnetic field power spectrum and is peaked in the squeezed configuration for a 
scale-invariant magnetic spectrum. 
By using the large angular part of the bispectrum generated by magnetic fields, 
the present bounds on non-Gaussianity set
a limit on the amplitude of the primordial magnetic field of the order of ${\cal O}(10)$~nGauss for the scale-invariant case and ${\cal O}(20)$~nGauss for the other spectral indexes.

\end{abstract}

\maketitle

\section{Introduction}

Cosmological inflation 
\cite{lrreview} has become the dominant paradigm to 
understand the initial conditions for the Cosmic Microwave Background (CMB) anisotropies
and structure formation. 
This picture has recently received further spectacular confirmation 
by the  Wilkinson Microwave 
Anisotropy Probe (WMAP) five year set of data \cite{wmap5}.
Present \cite{wmap5} and future \cite{planck} experiments 
may be sensitive to the non-linearities of the cosmological
perturbations at the level of second- or higher-order perturbation theory.
The detection of these non-linearities through the  non-Gaussianity
(NG) in the CMB \cite{review} has become one of the primary experimental targets. 

A possible source of NG could be primordial in 
origin, being specific to a particular mechanism for the generation of the cosmological perturbations. This is what 
makes a positive detection of NG so
relevant: it might help in discriminating among competing scenarios which otherwise might be indistinguishable. Indeed,
various models of inflation, firmly rooted in modern 
particle physics theory,   predict a significant amount of primordial
NG generated either during or immediately after inflation when the
comoving curvature perturbation becomes constant on super-horizon scales
\cite{review}. While single-field  \cite{noi}
and two(multi)-field \cite{two} models of inflation generically predict a tiny level of NG, 
`curvaton-type models',  in which
a  significant contribution to the curvature perturbation is generated after
the end of slow-roll inflation by the perturbation in a field which has
a negligible effect on inflation, may predict a high level of NG \cite{luw,ngcurv}.
Alternatives to the curvaton model are those models 
characterised by the curvature perturbation being 
generated by an inhomogeneity in 
the decay rate \cite{hk,varcoupling} or 
  the mass   \cite{varmass} or 
of the particles responsible for the reheating after inflation. 
Other opportunities for generating the curvature perturbation occur
 at the end of inflation \cite{endinflation} and  during
preheating \cite{preheating}.
All these models generate a level of NG which is local as the NG part of the primordial curvature
perturbation is a local function of the Gaussian part, being generated on superhorizon scales. 
In momentum space, the three point function, or bispectrum, arising from the local NG is dominated by the
so-called ``squeezed'' configuration, where one of the momenta is much smaller than the other two and it is parametrized by
the non-linearity parameter $f_{\rm NL}^{\rm loc}$. Other models, such as DBI inflation
\cite{DBI} and ghost inflation \cite{ghost}, predict a different kind of primordial
NG, called ``equilateral'', because the three-point function for this kind of NG is peaked on equilateral configurations, 
in which the lengths 
of the three wave-vectors forming a triangle in Fourier
space are equal \cite{Shapes}. The equilateral NG is parametrized by an amplitude $f_{\rm NL}^{\rm equil}$~\cite{CN}. Present limits
on NG are summarised by $-9<f^{\rm loc}_{\rm NL}<111$ and   $-151<f^{\rm equil}_{\rm NL}<253$ at 95\% CL \cite{wmap5,Curto}.

On the other hand there might exist other  sources of primordial NG in the CMB anisotropies beyond the primordial ones related
to the dynamics of inflation. One interesting possibility is the contribution to the non-Gaussian signal in the CMB anisotropies from 
a stochastic background of primordial magnetic fields. 
Large scale magnetic fields are almost everywhere in the universe,
from galaxies up to those present in galaxy clusters and
in the inter-cluster medium \cite{GR}.
The dynamo effect provides a
mechanism to explain the observed magnetic fields associated to galaxies,
whereas those associated to clusters may be generated
by gravitational compression. Both these mechanisms require
an initial magnetic seed, although with different amplitude and different 
correlation length.

Possible explanations for this initial seed have driven the interest in
primordial magnetic fields generated in the early universe. 
A stochastic background of primordial magnetic fields (PMFs) generated 
in the early universe with a mean 
amplitude well below micro-Gauss level can leave imprints on the temperature 
and polarisation anisotropy pattern of the cosmic microwave background (CMB).
The impact of a stochastic background of PMFs onto CMB anisotropies 
has distinctive imprints, such as a contribution in temperature 
which is larger than the CMB angular power spectrum sourced 
by scalar cosmological perturbation at high $\ell$ and a contribution in 
polarisation which include either $BB$ (generated by vector and tensor 
perturbations or by Faraday rotation \cite{faraday}) or parity-odd correlators as $TB$ 
(generated by an helical component \cite{Caprini:2003vc}).

As we mentioned, another distinctive imprint of PMF in CMB anisotropies is its 
non-Gaussian nature. The CMB signature of this type which has been first considered in the literature is due to a homogeneous PMF. A homogeneous magnetic field with fixed direction breaks spatial isotropy in the universe, and therefore leads to non-zero correlations between multipoles at different $\ell$, $\vev{a_{\ell-1,m}a_{\ell+1,m}}\neq 0$ \cite{CMBHom}. This effect has been first proposed in \cite{AlfvenHom}, and arises through the generation of vector metric pertubations from the Alfv\'{e}n waves magnetically induced in the primordial fluid. Recently, it has been reanalysed and found to reproduce, for a sufficiently high magnetic field amplitude, some of the anomalies of the CMB large scale fluctuations observed by WMAP such as the north-south asymmetry and the quadrupole-octopole alignment \cite{AnomalHom}. 

This effect is related to the presence of a homogeneous magnetic field (or equivalently, a stochastic magnetic field with correlation length larger than the horizon today). On the other hand, the CMB contribution of a stochastic background of PMFs, modelled as a fully inhomogeneous component, is intrinsically non-Gaussian: 
the PMFs energy-momentum tensor, the Lorentz force acting on baryons 
are quadratic in the magnetic field ${\bf B}(\eta,{\bf x})$, which 
is randomly distributed with a Gaussian distribution function.
The source terms to the Einstein-Boltzmann system are therefore  
$\chi$-distributed, leading to a PMF contribution to CMB 
fully non-Gaussian. Higher order statistical moments of the energy-momentum 
tensor of PMFs are therefore non-vanishing at leading order and are calculable 
with minimal assumptions, such as cutting sharply the power spectrum beyond a 
certain scale $k_D$ \cite{Brown:2005kr}. 

Non-gaussianities from PMFs are much less studied than those generated 
in inflationary methods. The study of the three point statistics of the 
PMF energy-momentum tensor in \cite{Brown:2005kr} 
is limited to the simplest particular collinear configuration. Nevertheless, 
due to the presence of a contribution from the collinear configuration, one deduces that 
non-gaussianities from PMFs can be different from the inflationary 
case, in which the collinear contribution is generically negligible 
with respect to the equilateral and squeezed ones (see however \cite{Holman:2007na}). 
In this paper we focus on the three point
statistics of the PMF energy density,  studying the contribution of all 
three configurations. 
By using the large scale relation between temperature 
anisotropies and PMF energy density given in Ref. \cite{Finelli:2008xh}, 
we  compute the 
temperature bispectrum and compare its contribution to the non-Gaussian statistics in the CMB anisotropies
with the present observational bounds.  
We also compare our results with those of the very recent paper \cite{Seshadri:2009sy}.

Our paper is organised as follows. In section II we introduce the stochastic 
background of primordial magnetic fields and discuss the infrared behaviour 
of the spectra of its energy-density. In section III and IV we discuss the 
CMB temperature spectrum and bispectrum induced by PMF on large scales. 
Section V is devoted to the analytic computation of the magnetic 
energy density bispectrum $\vev{\rho_B(\bk)\rho_B(\bq)\rho_B(\bp)}$ 
in general and for the collinear, squeezed and equilateral configurations.
In Section VI we insert these results into the CMB temperature bispectrum 
on large scales, and we give an estimation of the signal in section VII. 
In the first Appendix we derive analytic approximations to some integrals of Bessel functions which are useful to calculate both the spectrum and the bispectrum, and in the second Appendix we give the details for the exact computation of the energy density bispectrum in the collinear case for $n = 2, -2$.

\section{Primordial stochastic magnetic field}
\label{section2}

We adopt notations consistent with \cite{Kahniashvili:2006hy,Hu:1997hp}:
\be
B_i(\bx)=\int \frac{d^3k}{(2\pi)^3}e^{-i\bk\cdot\bx}B_i(\bk)~\rightarrow~
\de(\bk)=\int\frac{d^3x}{(2\pi)^3} e^{i\bk\cdot\bx} \nonumber\,,
\ee
where the definition of the delta function comes from $\int d^3k/(2\pi)^3\,e^{-i\bk\cdot\bx}\de(\bk)=1/(2\pi)^3$. With these conventions, the magnetic field power spectrum (defined as the Fourier transform of the two point correlation function) is\footnote{In this paper we neglect the possible presence of an helical component for the magnetic field, see for example \cite{Caprini:2003vc}}:
\bea
\vev{B_i(\bk)B^*_j(\bq)}&=&(2\pi)^3\de^3(\bk-\bq)(\de_{ij}-\hat{k}_i\hat{k}_j)P_B(k) \label{mpowerspectrum}\\
P_B(k)&=&A\,k^n\,,~~~k\leq k_D\,, \label{PB}
\eea
where $\hat{k}_i=k_i/k$, $A$ is a normalisation constant, $n$ the spectral index and $k_D$ the upper cutoff. Using the above equations we can define the mean square of the magnetic field as 
\be
\vev{B^2(\bx)}=\frac{A}{\pi^2}\frac{k_D^{n+3}}{n+3}\,.
\label{mean-squared}
\ee
If we are interested in the mean amplitude of the magnetic field on a given characteristic scale, we smooth the power spectrum over the chosen scale using a Gaussian filter: we have then $\vev{B^2(\bx)}|_\la=B_\la^2$ with 
\be
B_\la^2=\frac{1}{\pi^2}\int dk\,k^2\,P_B(k)\,e^{-k^2\la^2}=\frac{A}{2\pi^2}\frac{\Gamma[(n+3)/2]}{\la^{n+3}}\,,
\ee 
so that
\be
B_\la^2=\frac{\vev{B^2}}{2}\frac{n+3}{(k_D\la)^{n+3}}\Gamma{\big(\frac{n+3}{2}\big)}\,.
\label{Bla}
\ee
We also define the adimensional quantity $\Omega_B^{\rm tot}$ given by the ratio of the magnetic and the total radiation energy densities:
\be
\Om_B^{\rm tot}=\frac{\vev{B^2}}{8 \pi \rho_{\rm rel}}\simeq 10^{-7}\frac{\vev{B^2(\bx)}}{(10^{-9}\Gauss)^2}\,,
\label{Omtot}
\ee
where for the last equality we have used $\rho_{\rm rel}(\eta_0)\simeq 2\times 10^{-51}$ GeV$^4$, 
and $\eta_0$ denotes the conformal time today.

The upper cutoff $k_D$ corresponds to the damping scale, representing the dissipation of magnetic energy due to the 
generation of magneto-hydrodynamic waves \cite{Jedamzik:1996wp,Subramanian:1997gi}. Alfv\'{e}n waves are the most effective in dissipating 
magnetic energy, and in \cite{Subramanian:1997gi} it is demonstrated that around recombination the damping occurs at scales 
$k^{-1}\leqsim k^{-1}_D\simeq V_A L_{\rm Silk}$, where $V_A$ is the Alfv\'{e}n speed and $L_{\rm Silk}$ the Silk damping 
scale at recombination. Strictly speaking, Alfv\'{e}n waves are oscillatory perturbations superimposed on a homogeneous 
magnetic component, and the Alfv\'{e}n speed depends on the amplitude of the homogeneous component. In the cosmological context where the magnetic field is purely stochastic, the amplitude of this component can be taken as the one of a `low frequency' component obtained by smoothing the magnetic field amplitude over 
a sufficiently large scale \cite{Durrer:1999bk}. This scale corresponds to the Alfv\'{e}n scale at recombination, 
$k^{-1}_A\simeq V_A\eta_{\rm rec}$: magnetic modes on lager scales, in fact, do not have time to oscillate before 
recombination \cite{Subramanian:1997gi}. One has therefore $k_D/k_A=\eta_{\rm rec}/L_{\rm Silk}\simeq 10$, 
and $V_A^2=B_{L_A}^2/(4\pi(\rho+p))$. Consequently, the upper cutoff at the epoch of recombination is given by
\be
k_D\simeq \frac{1}{L_{\rm Silk}}\sqrt{\frac{16\pi}{3}\frac{\rho_{\rm rel}}{B_{L_A}^2}}=\frac{1}{L_{\rm Silk}}\frac{1}{\sqrt{\Om_B^{\rm tot}}}\sqrt{\frac{(2\pi)^{n+3}}{(n+3)\Gamma((n+3)/2)}}\left(\frac{k_A}{k_D}\right)^{\frac{n+3}{2}} \,.\label{kD}
\ee

Finiteness of the total magnetic energy density implies $n>-3$. In the rest of the paper we keep the spectral index as a  free parameter, when possible; however, in order to carry on our calculations analytically we sometimes need to specify it. For example, in Section~\ref{sec:bispectrumrho}, we choose the values $n=2$ and $n=-2$. $n=2$ is the lowest possible value for a magnetic field generated by a causal process \cite{Durrer:2003ja}, such as a phase transition \cite{phase}, or a charge separation process operating around recombination \cite{charge}. A magnetic field generated during inflation \cite{inflation} (or by any other a-causal process such as, for example, in pre big bang theories \cite{prebig}), can take any value of $n>-3$. However, because of Nucleosynthesis constraints \cite{Caprini:2001nb}, only for very red spectra the magnetic field can assume sufficiently high amplitudes to have an impact in the CMB. Therefore, in the following we choose the value $n=-2$ (for some analytic calculations), or $n\rightarrow -3$ when possible. 

The spatial part of the energy momentum tensor of the magnetic field is
\be
\tau_{ij}(\bx)=\frac{1}{4\pi}\left[\frac{1}{2}\de_{ij}B_l(\bx)B_l(\bx)-B_i(\bx)B_j(\bx)\right]\,,
\ee
and the magnetic energy density
\bea
\rho_B(\bx)&=&\tau_{ii}(\bx)=\frac{1}{8\pi}B^2(\bx)\,,\label{rhoBx}\\
\rho_B(\bk)&=&\frac{1}{8\pi}\int\frac{d^3p}{(2\pi)^3}B_i(\bk-\bp)B_i(\bp)\label{rhoBk}\,.
\eea
As we will see in the next section, to calculate the CMB temperature spectrum from the Sachs Wolfe effect we need the power spectrum of the magnetic energy density:
\be
\vev{\rho_B(\bk)\rho_B^*(\bq)}\equiv (2\pi)^3\de(\bk-\bq)|\rho_B(k)|^2 =\frac{2}{(8\pi)^2}\de(\bk-\bq)\int d^3p\,P_B(\bp)P_B(|\bk-\bp|)(1+\mu^2)\,,
\ee
the second equality is obtained using \eqref{rhoBk}, and $\mu=\hat{p}\cdot\widehat{\bk-\bp}$. Therefore
\be
|\rho_B(k)|^2=\frac{1}{256\pi^5}\int d^3p\,P_B(\bp)P_B(|\bk-\bp|)(1+\mu^2)\,.
\label{specrhoB}
\ee
As demonstrated in Ref. \cite{Finelli:2008xh}, $|\rho_B(k)|^2$ goes to zero at a wavenumber corresponding to twice the magnetic field spectrum cutoff, $k=2k_D$. Eq.~(21) of \cite{Finelli:2008xh} gives the behaviour of $|\rho_B(k)|^2$ at large scales $k\ll k_D$ and for spectral indexes $n>-3/2$: the generic behaviour in this case 
is white noise \cite{Finelli:2008xh}
\be
|\rho_B(k)|^2\simeq \frac{A^2\,k_D^{2n+3}}{32\pi^4(2n+3)}\label{rhoBwhitenoise}\,.
\ee
For $n=-2$, an exact calculation as in Refs. \cite{Finelli:2008xh,PFP} gives the behaviour for $|\rho_B(k)|^2$ at large scales $k\ll k_D$ as
\be
|\rho_B(k)|^2\simeq \frac{3 A^2}{512 \, \pi^2 \, k} \,,
\label{rhoBnmenodue}
\ee
For $n<-3/2$ we use the approximated formula given by Ref.~\cite{Kahniashvili:2006hy}:
\be
|\rho_B(k)|^2\simeq \frac{3A^2}{128\pi^4} \frac{n}{(2n+3)(n+3)} k^{2n+3}\label{rhoBmen3mezzi}\,.
\ee
For $n=-2$, expressions (\ref{rhoBnmenodue}) and (\ref{rhoBmen3mezzi}) are in agreement concerning 
the dependence on the wavenumber, but the numerical factor differs by a factor $\pi^2/8$.

\section{CMB temperature spectrum at large angular scales}
\label{sec:sachswolfe}

We use the characterisation of the CMB temperature anisotropy induced by a stochastic background of primordial magnetic 
fields given in \cite{PFP}.
It is useful to define the adimensional magnetic energy parameter in $k$-space
\be
\Om_B(\bk)=\frac{\rho_B(\bk)}{\rho_{\rm rel}}\label{OmB}\,.
\ee
From the initial conditions given in \cite{Finelli:2008xh,PFP} we parametrize the temperature anisotropy as  
\be
\frac{1}{4}\de_\gamma+\psi=\frac{\al}{4}\Om_B(\bk)\,,
\ee
where $\al\sim 0.1$ is a multiplication constant required since the above equation would be exact with $\alpha=1$ in the radiation era.
Therefore, the temperature anisotropy is given in terms of this quantity as
\be
\frac{\Theta_\ell^{(0)}(\eta_0,\bk)}{2\ell+1}=\frac{\al}{4}\Om_B(\bk)j_\ell(k(\eta_0-\eta_{\rm dec}))\,,
\label{thetaSW}
\ee
where $j_\ell$ is the spherical Bessel function and $\eta_0$, $\eta_{\rm dec}$ denote conformal time respectively today and at decoupling. 
The CMB power spectrum is therefore \cite{Hu:1997hp}:
\be
C_\ell^B=\frac{2}{\pi}\int_0^\infty dk\,k^2\,\frac{\vev{\Theta_\ell^{(0)}(\eta_0,\bk)\Theta_\ell^{(0)*}(\eta_0,\bk)}}{(2\ell+1)^2}=\frac{\al^2}{8\pi}\int_0^\infty dk\,k^2\,|\Om_B(\bk)|^2j_\ell^2(k(\eta_0-\eta_{\rm dec}))\,.
\label{Clspectrum}
\ee
In the case $n>-3/2$, substituting definition 
(\ref{OmB}) and \eqref{rhoBwhitenoise} in the above equation we have \cite{PFP}
\be
C_\ell^B\simeq\frac{\al^2\,A^2\,k_D^{2n+6}}{8(2\pi)^5(2n+3)\rho_{\rm rel}^2\,(k_D\eta_0)^3}\int_0^{k_D\eta_0} dx\,x^2\,j^2_\ell(x)\simeq \frac{\al^2}{512\pi}\frac{(n+3)^2}{2n+3}\frac{\vev{B^2}^2}{\rho_{\rm rel}^2}\frac{1}{(k_D\eta_0)^2}~~~~{\rm for}~n>-3/2\,,
\label{CMBspecSachsWolfe}
\ee
where $x=k\eta_0$, we have approximated $j_\ell(k(\eta_0-\eta_{\rm dec}))\simeq j_\ell(k\eta_0)$ and we integrate only up to the upper cutoff $k_D$ since we are using the approximated expression \eqref{rhoBwhitenoise} which is strictly valid only for $k\ll k_D$. For the second equality in the above equation, we have approximated the integral as given in \eqref{intspectrum} of appendix \ref{appendixAA}, since we have that $y=k_D\eta_0\gg 1$. We have also used \eqref{mean-squared} to express the result in terms of the mean squared magnetic field.

For $n=-2$, we use instead \eqref{rhoBnmenodue}: substituting it in \eqref{Clspectrum}, we find
\be
C_\ell^B\simeq\frac{3\,\al^2\,A^2\,k_D^{2}}{4096\,\pi^3\rho_{\rm rel}^2\,(k_D\eta_0)^2}\int_0^{k_D\eta_0} dx\,x\,j^2_\ell(x)\simeq \frac{3\,\pi\,\al^2}{8192}\frac{\vev{B^2}^2}{\rho_{\rm rel}^2}\frac{1}{(k_D\eta_0)^2}\,\log\left(\frac{k_D\eta_0}{\ell}\right)~~~~{\rm for}~n=-2\,,
\label{CMBspecSachsWolfenmenodue}
\ee
where in the second equality we use the approximation given in \eqref{approx1} of appendix \ref{appendixAA}. For more negative values of $n$, $n<-2$, in the absence of an exact expression, we use \eqref{rhoBmen3mezzi}: the CMB spectrum becomes
\bea
C_\ell^B&\simeq&\frac{3\,\al^2\,A^2\,k_D^{2n+6}}{1024\,\pi^5\rho_{\rm rel}^2\,(k_D\eta_0)^{2n+6}}\frac{n}{(2n+3)(n+3)}\int_0^{k_D\eta_0} dx\,x^{2n+5}\,j^2_\ell(x)\nonumber \\
&\simeq& \frac{3\,\al^2}{4096\,\sqrt{\pi}}\frac{n(n+3)}{(2n+3)}\frac{\Gamma[-n-2]}{\Gamma[-n-3/2]}\frac{\vev{B^2}^2}{\rho_{\rm rel}^2}\frac{1}{(k_D\eta_0)^{2n+6}}\ell^{2n+4}~~~~{\rm for}~n<-2\,,
\label{specnmen3mezzi}
\eea
where again for the second equality we have used \eqref{approxmin1} of appendix \ref{appendixAA}. 
The slope in $\ell$ of this last expression is only approximatively recovered numerically, whereas it is perfectly recovered 
for Eqs. (\ref{CMBspecSachsWolfe},\ref{CMBspecSachsWolfenmenodue}).

\section{CMB temperature bispectrum at large angular scales}

We want to evaluate the CMB angular bispectrum of the temperature anisotropy due to the Sachs Wolfe effect induced by the magnetic field energy density. The angular bispectrum is given by $\vev{a_{\ell_1m_1}a_{\ell_2m_2}a_{\ell_3m_3}}$, with the spherical harmonic expansion coefficients 
\be
a_{\ell m}(\bx)=\int d\Om_{\hat{n}}Y^*_{\ell m}(\hat{n};\hat{e})\Theta^{(0)}(\bx,\hat{n}) \label{alm}\,,
\ee
where $Y^*_{\ell m}(\hat{n};\hat{e})$ is the spherical harmonic with respect to a basis where $\hat{e}$ is an arbitrary but fixed direction, and $\Theta^{(0)}(\bx,\hat{n})$ is the scalar temperature perturbation at position $\bx$ ($\hat{n}$ is the direction of light propagation). Using the formalism developed in \cite{Hu:1997hp} one has
\bea
\Theta^{(0)}(\bx,\hat{n})=\int \frac{d^3k}{(2\pi)^3}\Sigma_\ell \Theta^{(0)}_\ell(\eta_0,\bk) G^0_{\ell}\,,\\
G^0_\ell=(-i)^\ell\sqrt{\frac{4\pi}{2\ell+1}}Y_{\ell 0}(\hat{n};\hat{k}) e^{i\bk\cdot\bx}\,,
\eea
with respect to a basis where $\hat{k}$ is fixed. Substituting the above expressions in \eqref{alm}, and changing basis accordingly (\emph{cf.} \cite{libroruth}), one finds 
\be
a_{\ell m}(\bx)=\frac{4\pi(-i)^\ell}{2\ell+1}\int \frac{d^3k}{(2\pi)^3} \Theta^{(0)}_\ell(\eta_0,\bk) e^{i\bk\cdot\bx}Y^*_{\ell m}(\hat{k};\hat{e})\,.
\ee
Therefore the angular bispectrum is given by (we place the observer in $\bx=0$)
\bea
\vev{a_{\ell_1m_1}a_{\ell_2m_2}a_{\ell_3m_3}}&=&\frac{(4\pi)^3(-i)^{\ell_1+\ell_2+\ell_3}}{(2\ell_1+1)(2\ell_2+1)(2\ell_3+1)}\int \frac{d^3k\,d^3q\,d^3p}{(2\pi)^9}\,Y^*_{\ell_1 m_1}(\hat{k};\hat{e})Y^*_{\ell_2 m_2}(\hat{q};\hat{e})Y^*_{\ell_3 m_3}(\hat{p};\hat{e}) \label{aaa}\\
&\times&\vev{\Theta^{(0)}_{\ell_1}(\eta_0,\bk)\Theta^{(0)}_{\ell_2}(\eta_0,\bq)\Theta^{(0)}_{\ell_3}(\eta_0,\bp)}\,.\nonumber
\eea
Remembering \eqref{thetaSW} and definition (\ref{OmB}), we see that in order to proceed we need to evaluate the bispectrum of the magnetic energy density $\vev{\rho_B(\bk)\rho_B(\bq)\rho_B(\bp)}$. 

\section{The magnetic energy density bispectrum}
\label{sec:bispectrumrho}

From the expression of the magnetic energy density given in \eqref{rhoBk}, we see that its bispectrum is given in terms of the six point correlation function of the magnetic field
\be
\vev{\rho_B(\bk)\rho_B(\bq)\rho_B(\bp)}=\frac{1}{(8\pi)^3}\int\frac{d^3\tilde{k}\,d^3\tilde{q}\,d^3\tilde{p}}{(2\pi)^9}
\vev{B_i(\tilde{\bk})B_i(\bk-\tilde{\bk})B_j(\tilde{\bq})B_j(\bq-\tilde{\bq})B_l(\tilde{\bp})B_l(\bp-\tilde{\bp})}\,.
\label{rhoBrhoBrhoB}
\ee
Since the magnetic field is assumed to be a Gaussian variable, we can use Wick's theorem to decompose the six point correlation function into products of the magnetic field power spectrum. To compute the above expression, we then use the definition of the magnetic power spectrum \eqref{mpowerspectrum} and the fact that $B_i^*(\bk)=B_i(-\bk)$. Of the total fifteen terms obtained using Wick's theorem, seven are irrelevant because they are one-point terms proportional to $\de(\bk)$, $\de(\bq)$ or $\de(\bp)$, while eight terms survive. Each of the latter is the product of three power spectra, and consequently contains the product of three delta functions (\cf \eqref{mpowerspectrum}). Two delta functions can be integrated, while the remaining one reduces to $\de(\bk+\bq+\bp)$, the homogeneity condition. Starting from \eqref{rhoBrhoBrhoB}, the final result depends on which of the variables of the triple integral remains. For example, performing the integration in $d^3\tilde{p}$ and
  $d^3\tilde{q}$ and leaving out $d^3\tilde{k}$, one obtains (appropriately renaming the mute indexes)
\be
\vev{\rho_B(\bk)\rho_B(\bq)\rho_B(\bp)}=\frac{1}{128\pi^3}\de(\bk+\bp+\bq)\int d^3\tilde{k}\,
P_{ij}(\tilde{\bk})P_{jl}(\bk-\tilde{\bk})[P_{il}(\bq+\tilde{\bk})+P_{il}(\bp+\tilde{\bk})]\,,
\ee
where for conciseness we have defined
\be
P_{ij}(\bk)=P_B(k)\,(\de_{ij}-\hat{k}_i\hat{k}_j)
\label{indexspectrum}
\ee
(note that $P_{ij}(\bk)=P_{ij}(-\bk)$). On the other hand, integrating out $d^3\tilde{k}$ and $d^3\tilde{p}$ one obtains
\be
\vev{\rho_B(\bk)\rho_B(\bq)\rho_B(\bp)}=\frac{1}{128\pi^3}\de(\bk+\bp+\bq)\int d^3\tilde{q}\,
P_{ij}(\tilde{\bq})P_{jl}(\bq-\tilde{\bq})[P_{il}(\bk+\tilde{\bq})+P_{il}(\bp+\tilde{\bq})]\,,
\ee
while integrating out $d^3\tilde{q}$ and $d^3\tilde{k}$ one obtains
\be
\vev{\rho_B(\bk)\rho_B(\bq)\rho_B(\bp)}=\frac{1}{128\pi^3}\de(\bk+\bp+\bq)\int d^3\tilde{p}\,
P_{ij}(\tilde{\bp})P_{jl}(\bp-\tilde{\bp})[P_{il}(\bk+\tilde{\bp})+P_{il}(\bq+\tilde{\bp})]\,.
\ee
This is just a consequence of the fact that the right hand side of \eqref{rhoBrhoBrhoB} is not apparently symmetric under the exchange of $\bk$, $\bq$ and $\bp$, contrary to the left hand side. Since the final result should be symmetric, we finally set:
\bea
\vev{\rho_B(\bk)\rho_B(\bq)\rho_B(\bp)}=\frac{\de(\bk+\bp+\bq)}{384\pi^3}&&\left\{
\int d^3\tilde{k}\,P_{ij}(\tilde{\bk})P_{jl}(\bk-\tilde{\bk})[P_{il}(\bq+\tilde{\bk})+P_{il}(\bp+\tilde{\bk})]\right. \nonumber \\
&&+\int d^3\tilde{k}\, P_{ij}(\tilde{\bk})P_{jl}(\bq-\tilde{\bk})[P_{il}(\bk+\tilde{\bk})+
P_{il}(\bp+\tilde{\bk})]\nonumber\\
&&+\left.\int d^3\tilde{k}\, P_{ij}(\tilde{\bk})P_{jl}(\bp-\tilde{\bk})[P_{il}(\bq+\tilde{\bk})+
P_{il}(\bk+\tilde{\bk})]\right\}\,.
\label{rhoBsymm}
\eea
Using definitions (\ref{indexspectrum}) and (\ref{PB}) we have in all generality:
\bea
P_{ij}(\bk)P_{jl}(\bq)P_{il}(\bp)=A^3 k^n p^n q^n [(\hat{k}\cdot\hat{q})^2+(\hat{k}\cdot\hat{p})^2+(\hat{q}\cdot\hat{p})^2-(\hat{k}\cdot\hat{q})(\hat{k}\cdot\hat{p})(\hat{q}\cdot\hat{p})]
~~{\rm if}~k\leq k_D\,,~q\leq k_D\,,~p\leq k_D\,, \label{productP}
\eea
and zero else. Due to the complexity of the angular structure and of the integration boundary of the integrals in \eqref{rhoBsymm}, we cannot derive an exact expression for $\vev{\rho_B(\bk)\rho_B(\bq)\rho_B(\bp)}$ which is valid for any configuration of $\bk,~\bq,~\bp$. We can however give an analytical estimate of the result, which we present in the following.

\begin{figure}
\includegraphics[height=7cm,width=9cm]{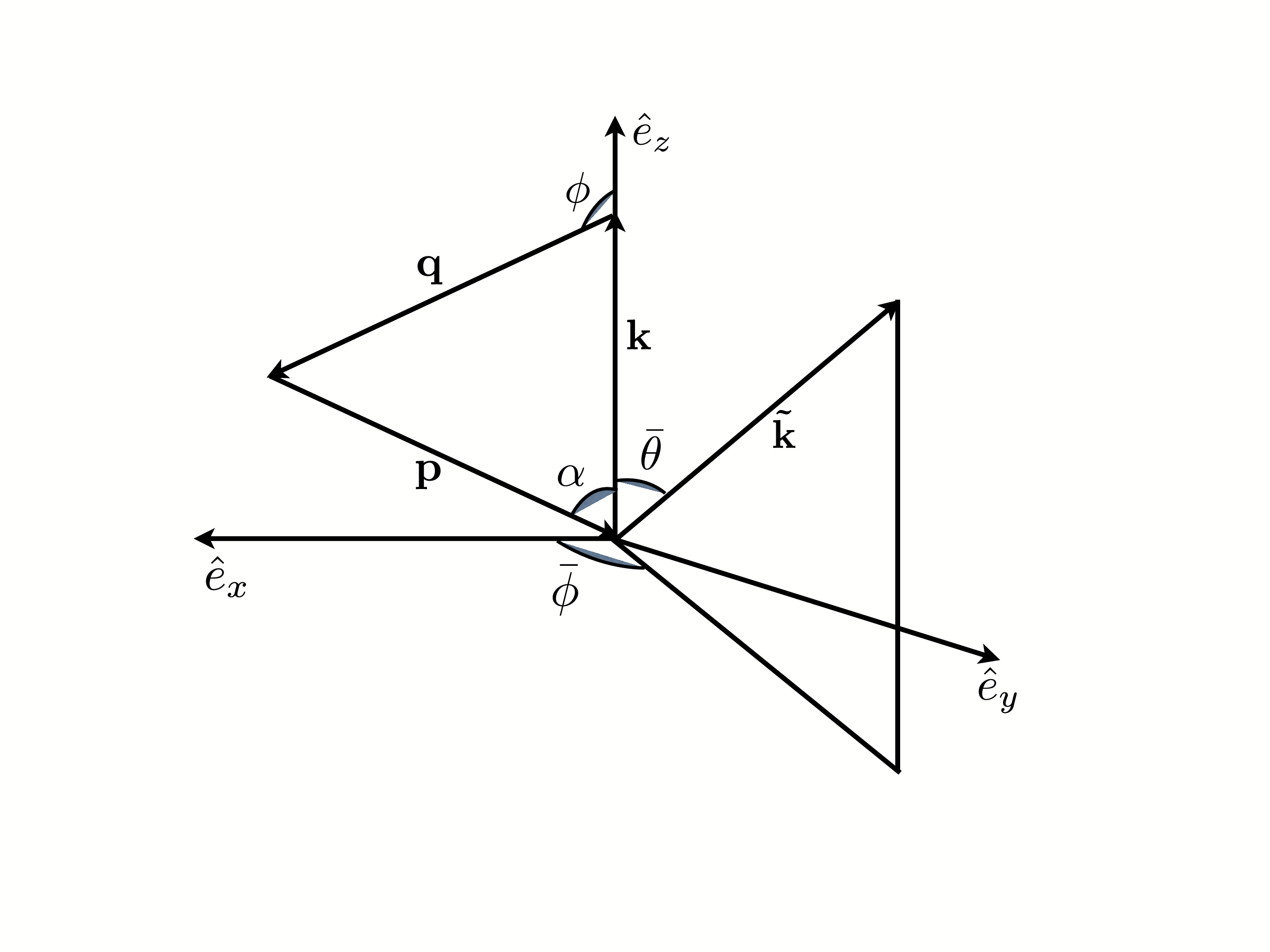}
\caption{The geometrical configuration used to perform the integration: $\bk,~\bq$ and $\bp$ are free, while ${\bf\tilde{k}}$ is the integration wave-vector (see \cite{Brown:2005kr}).}
\label{geometry}
\end{figure}

We are interested in estimating the behaviour of the integrals in (\ref{rhoBsymm}). From the expression in Eq.~\ref{productP} it is clear that, depending on the value of the spectral index $n$, the integral could diverge in the infrared limit. On the other hand, the angular part always gives a finite contribution. We therefore neglect the angular part for the following estimate, and set
\be
\vev{\rho_B(\bk)\rho_B(\bq)\rho_B(\bp)}\simeq \frac{\de(\bk+\bp+\bq)}{384\pi^3}A^3\left\{ \int d^3\tilde{k} \,\tilde{k}^n |\bk-\tilde{\bk}|^n \left[ |\bq+\tilde{\bk}|^n+|\bp+\tilde{\bk}|^n\right] + {\rm permutations}\right\}\,.
\label{rhoBapp}
\ee 
To perform the above integration, following \cite{Brown:2005kr}, we choose a basis with $\hat{e}_z\parallel\bk$ and where the triangle formed by $\bk,~\bq,~\bp$ lies in the plane perpendicular to $\hat{e}_y$, in $y=0$, see Fig.~\ref{geometry}. We call $\phi$ the angle between $\bk$ and $\bq$, $\cos \phi=\hat{k}\cdot\hat{q}$, and $\alpha$ the angle between $\bk$ and $-\bp$, $\cos (\pi-\alpha)=\hat{k}\cdot\hat{p}$. The integration variable $\tilde{\bk}$ has angles $\bar{\theta}$ with $\hat{e}_z\parallel\bk$ and $\bar{\phi}$ with the plane identified by the triangle formed by $\bk,~\bq,~\bp$ (\emph{cf.} Fig.~\ref{geometry}). The angle between $\tilde{\bk}$ and $\bq$ is expressed in terms of the previously defined ones as 
\be
\hat{\tilde{k}}\cdot\hat{q}=\sin\bar{\theta}\cos\bar{\phi}\sin\phi+\cos\bar{\theta}\cos\phi\,,
\ee
and the one between $\tilde{\bk}$ and $\bp$ is
\be
\hat{\tilde{k}}\cdot\hat{p}=-(\sin\bar{\theta}\cos\bar{\phi}\sin\alpha+\cos\bar{\theta}\cos\alpha)\,.
\ee
We remind that the boundaries of the integrals in (\ref{rhoBapp}) are defined by the condition that the momenta coming from the power spectrum are bounded by $k_D$: in the first integral for example, the conditions are $\tilde{k}\leq k_D$, $|\bk-\tilde{\bk}|\leq k_D$, $|\bq+\tilde{\bk}|\leq k_D$. 

Let us first concentrate on the first integral of (\ref{rhoBapp}). For negative values of $n$, it has integrable divergences for $\tilde{\bk}\rightarrow \bk$ and for $\tilde{\bk}\rightarrow -\bq$. We approximate the total result by selecting only these angular configurations, which are the biggest contributions to the integral for negative $n$, and are at least representative of the total result for positive $n$. By doing so, and using the above reference system, the first integral of (\ref{rhoBapp}) becomes ($\bar{\Omega}$ denotes the angular boundary)
\bea
\int d^3\tilde{k} \,\tilde{k}^n |\bk-\tilde{\bk}|^n |\bq+\tilde{\bk}|^n &=& \int_0^{k_D} d\tilde{k} \,\tilde{k}^{n+2} \int_{\bar{\Omega}} d\Omega \, \left[k^2+\tilde{k}^2-2k\tilde{k}\cos\bar\theta\right]^{n/2}\left[q^2+\tilde{k}^2+2q\tk (\sin\bar{\theta}\cos\bar{\phi}\sin\phi+\cos\bar{\theta}\cos\phi) \right]^{n/2} \nonumber \\
&\simeq& 2\pi \int_0^{k_D} d\tilde{k} \,\tilde{k}^{n+2}  \left[ |k-\tk|^n (q^2+\tk^2 +2q\tk \cos\phi)^{n/2} 
+ (k^2+\tk^2+2k\tk\cos\phi)^{n/2}|q-\tk|^n \right]
\eea
where in the second equality we have accounted only for the two above mentioned angular configurations: the first term of the second equality is the contribution of the angular configuration $\tilde{\bk}\rightarrow \bk$, and therefore $\bar\theta=0$; the second one is the contribution of the angular configuration $\tilde{\bk}\rightarrow -\bq$, and therefore $\bar\theta=\pi-\phi$ and $\bar\phi=\pi$. We have inserted the factor $2\pi$ to simulate the integration in $d\bar\phi$, which should be present at least in the first configuration. We repeat the same approximation scheme in each term of (\ref{rhoBapp}), to obtain finally
\bea
\label{rhoconfig}
\vev{\rho_B(\bk)\rho_B(\bq)\rho_B(\bp)}&\simeq& \frac{\de(\bk+\bp+\bq)}{96\pi^2}A^3\,\times \\
&& \left\{ \int_0^{k_D} d\tk\, \tk^{n+2} \left[ |k-\tk|^n (q^2+\tk^2 +2q\tk \cos\phi)^{n/2} + (k^2+\tk^2+2k\tk\cos\phi)^{n/2}|q-\tk|^n \right] \right. \nonumber \\
&& + \int_0^{k_D} d\tk\, \tk^{n+2} \left[ |k-\tk|^n (p^2+\tk^2 -2p\tk \cos\alpha)^{n/2} + (k^2+\tk^2-2k\tk\cos\alpha)^{n/2}|p-\tk|^n \right] \nonumber \\
&& + \left. \int_0^{k_D} d\tk\, \tk^{n+2} \left[ |q-\tk|^n (p^2+\tk^2 -2p\tk \cos(\phi-\alpha))^{n/2} + (q^2+\tk^2-2q\tk\cos(\phi-\alpha))^{n/2}|p-\tk|^n \right] \right\}\,. \nonumber
\eea
Note that the terms in \eqref{rhoBapp} which share the same wave-vectors collect two by two for the angular configurations considered (\emph{c.f.} \eqref{rhoBsymm}). It is now possible to evaluate approximatively the above integrals. As already mentioned, the apparent divergence for negative $n$ is integrable. Assuming $k < q < k_D$, we approximate the first integral in the above expression as  
\bea
\int_0^{k_D} d\tk\, \tk^{n+2} \left[ |k-\tk|^n (q^2+\tk^2 +2q\tk \cos\phi)^{n/2} + (k^2+\tk^2+2k\tk\cos\phi)^{n/2}|q-\tk|^n \right] \simeq \nonumber\\
2\left( q^n\,k^n\,\int_0^k d\tilde{k} \,\tilde{k}^{n+2} + q^n \int_k^q d\tilde{k} \,\tilde{k}^{2n+2} + \int_q^{k_D} d\tilde{k} \,\tilde{k}^{3n+2}\right)\,.
\eea
We see that, under this approximation, the angular part plays no longer a role, and the result is the same for the two terms of the first line of the above equation. 

Applying the same technique for each integral in \eqref{rhoconfig}, for the combination $k\leq q \leq p\leq k_D$ we find the total approximate behaviour:
\bea
\label{approxrhoB}
\vev{\rho_B(\bk)\rho_B(\bq)\rho_B(\bp)}&\simeq& \frac{\de(\bk+\bp+\bq)}{48\pi^2}A^3\,\times \\
&& \left\{ \frac{n}{(n+3)(2n+3)}q^n k^{2n+3} + \frac{n}{(3n+3)(2n+3)}q^{3n+3} +\frac{k_D^{3n+3}}{3n+3}\right.
\nonumber \\
&& +\frac{n}{(n+3)(2n+3)}p^n k^{2n+3} + \frac{n}{(3n+3)(2n+3)}p^{3n+3} +\frac{k_D^{3n+3}}{3n+3} \nonumber \\
&& \left. +\frac{n}{(n+3)(2n+3)}p^n q^{2n+3} + \frac{n}{(3n+3)(2n+3)}p^{3n+3} +\frac{k_D^{3n+3}}{3n+3}\right\}
~~~~~~~{\rm for}~k\leq q\leq p\leq k_D \,, \nonumber 
\eea
while if $q\leq k\leq p$ we have to exchange $k$ and $q$ in the above expression, and so on with all the ordered permutations of the wave-numbers. 

Observing \eqref{approxrhoB}, we can confirm what pointed out in \cite{Brown:2005kr}, \emph{i.e.} that there are two distinctive spectral regimes for the bispectrum. For flat and blue magnetic field spectra, with $n>-1$, the infrared limit $k\rightarrow 0$ of the bispectrum is white noise; \eqref{approxrhoB} is in fact dominated by the constant terms $k_D^{3n+3}/(3n+3)$. On the other hand, for red magnetic field spectra $n<-1$, the bispectrum is divergent in the infrared limit. As we will see in the next sections, the divergence can go as $k^{2n+3}$ or as $k^{3n+3}$, depending on the wave-vector configuration. The same behaviour holds for the magnetic energy density power spectrum, but in this case the discriminating value is $n=-3/2$, and the infrared divergence for $n<-3/2$ goes as $k^{2n+3}$ \cite{Kahniashvili:2006hy,Finelli:2008xh} (in \cite{Finelli:2008xh}, it has been found that for the limiting value $n=-3/2$ the white noise spectrum acquires a logarithmic dependence on $k$: this is the case also here for the corresponding limiting value $n=-1$). The above approximated result is valid only for $k,~q$ and $p$ smaller than the magnetic upper cutoff $k_D$, while in general they do not need to satisfy this bound. As already mentioned in section \ref{section2}, in \cite{Finelli:2008xh} it has been found that the magnetic energy spectrum goes to zero at $k=2k_D$, due to the convolution boundaries (\emph{c.f} \eqref{specrhoB}). As we will see in the next sections, the same behaviour holds also for the bispectrum (this is verified exactly in the collinear configuration).

The above equation (\ref{approxrhoB}) is a general approximation to the magnetic field energy density bispectrum in the infrared limit. We now compare it with the result coming from a specific configuration of the wave-vectors, the collinear configuration, for which we have an exact result. We find that the above expression can be considered quite a good approximation to the true magnetic field bispectrum in the infrared limit. We also give explicit formulas for the squeezed and equilateral configurations, for which, however, we do not calculate the exact result. We find that the three configurations give a comparable white noise contribution for $n>-1$, while if $n<-1$ the collinear and equilateral configurations diverge in the infrared limit as $k^{3n+3}$, while the squeezed one diverges as $k^{2n+3}$.

\subsection{Collinear configuration}

The collinear (or flattened) configuration is given by two equal wave-vectors, while the third one points in the opposite direction: for example, $\bp=\bq$ and $\bk=-2\bq$. In this case, it is possible to calculate the bispectrum (\ref{rhoBsymm}) exactly. The three permutations of $\bk,~\bq,~\bp$ of this configuration should be present in the symmetric expression (\ref{rhoBsymm}): this gives in the end
\bea
\left.\vev{\rho_B(\bk)\rho_B(\bq)\rho_B(\bp)}\right|_{\rm collinear}=
\frac{\de(\bk+\bp+\bq)}{384\pi^3}& & \frac{2}{3} \int d^3\tilde{k}\,P_{ij}(\tilde{\bk})\left\{  P_{jl}\left(\frac{\bk}{2}+\tilde{\bk}\right)
\left[ P_{il}(\bk+\tilde{\bk})+P_{il}\left(\frac{\bk}{2}-\tilde{\bk}\right)\right] \right. \nonumber \\
&+& \left.
P_{jl}(\bk-\tilde{\bk}) P_{il}\left(\frac{\bk}{2}-\tilde{\bk}\right) 
+ \bk \rightarrow \bp \, + \, \bk \rightarrow \bq \right\} \,.
\label{collinear}
\eea

Therefore, for the collinear case we find the following expression, using \eqref{productP}:
\bea\label{collinear2}
& &\left.\vev{\rho_B(\bk)\rho_B(\bq)\rho_B(\bp)}\right|_{\rm collinear}=
\frac{\de(\bk+\bp+\bq)}{576\pi^3}A^3  \times\nonumber\\
& & \left\{ 2 \int_{V_1} d^3\tilde{k} ~ \tilde{k}^n\,\left| \frac{\bk}{2}+\tilde{\bk} \right|^n \left|\bk+\tbk\right|^n 
\left[\frac{(\hat{\tilde{k}}\cdot \bk +2\,\tilde{k})^2}{4\,\left| \frac{\bk}{2}+\tilde{\bk} \right|^2}+
\frac{(\hat{\tilde{k}}\cdot \bk +\tilde{k})^2}{\left|\bk+\tilde{\bk} \right|^2}+
\frac{(k^2+3\,\tbk\cdot \bk +2\,\tilde{k}^2)(k^2-(\hat{\tilde{k}}\cdot \bk)^2)}{4\,\left| \frac{\bk}{2}+\tilde{\bk} \right|^2\left|\bk+\tilde{\bk} \right|^2}\right] \right. \nonumber \\
& & + \int_{V_2} d^3\tilde{k} ~ \tilde{k}^n\,\left| \frac{\bk}{2}+\tilde{\bk} \right|^n \left|\frac{\bk}{2}-\tbk\right|^n 
\left[\frac{(\hat{\tilde{k}}\cdot \bk+2\,\tilde{k})^2}{4\,\left| \frac{\bk}{2}+\tilde{\bk} \right|^2}+
\frac{(\hat{\tilde{k}}\cdot \bk -2\,\tilde{k})^2}{4\,\left|\frac{\bk}{2}-\tilde{\bk} \right|^2}+
\frac{(k^2-4\,\tilde{k}^2)(k^2-(\hat{\tilde{k}}\cdot \bk)^2)}{16\,\left| \frac{\bk}{2}+\tilde{\bk} \right|^2\left|\frac{\bk}{2}-\tilde{\bk} \right|^2}\right] \nonumber \\
& & + \left. \,\bk \rightarrow \bp \, + \, \bk \rightarrow \bq \right\} \,,
\eea
where $V_1$ denotes the volume given by the three conditions 
\bea
\tilde{k}&\leq & k_D\nonumber \\ 
|\bk/2+\tilde{\bk}|&\leq& k_D \nonumber \\
|\bk+\tilde{\bk}|&\leq& k_D\,,
\eea 
and $V_2$ is given by the conditions
\bea
\tilde{k}&\leq & k_D\nonumber \\ 
|\bk/2+\tilde{\bk}|&\leq& k_D \nonumber \\
|\bk/2-\tilde{\bk}|&\leq& k_D\,.
\eea 
The last term of Eq.~\ref{collinear}, $P_{ij}(\tilde{\bk})P_{jl}(\bk-\tilde{\bk}) P_{il}\left(\frac{\bk}{2}-\tilde{\bk}\right)$, becomes equal to the first one by changing $\tilde{\bk}$ to $-\tilde{\bk}$ and the integration volume accordingly. 

\begin{figure}
\includegraphics[height=4cm,width=5cm]{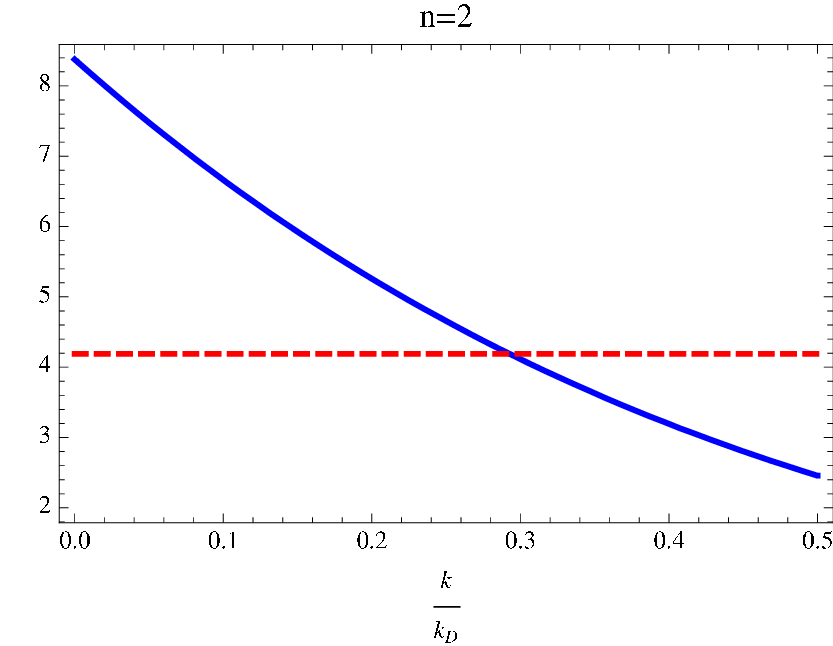}
\includegraphics[height=4.2cm,width=5cm]{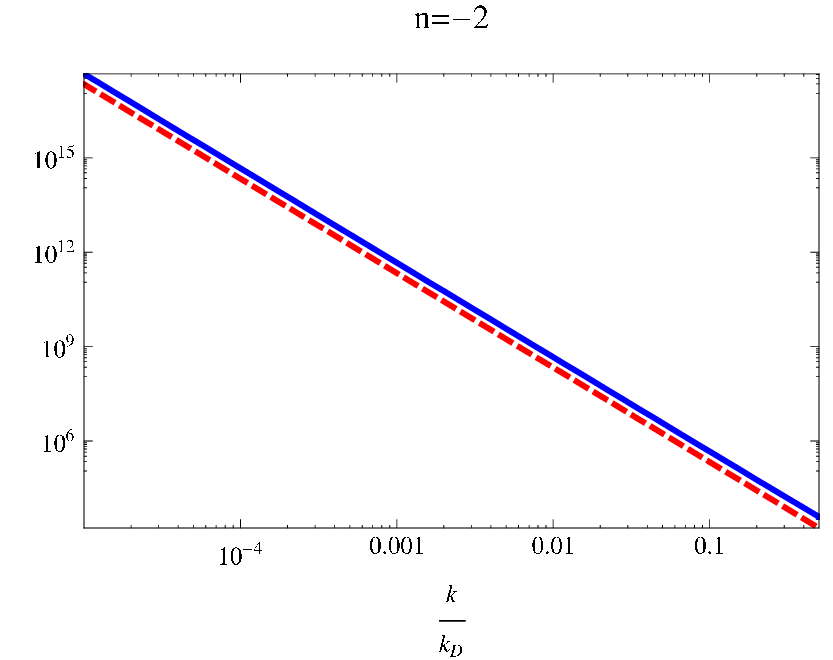}
\caption{The magnetic field bispectrum in the collinear configuration $\bp=\bq=-\bk/2$, normalised by the quantity $A^3 k_D^{3n+3} / (576\pi^3)$, as a function of $k/k_D$, for $n=2$ (left plot) and $n=-2$ (right plot). We only show the infrared region $k\leq k_D/2$. The blue, solid line is the exact result, while the red, dashed line the approximation given in \eqref{approxcoll}.}
\label{figcollinear}
\end{figure}

\begin{figure}
\includegraphics[height=4cm,width=5cm]{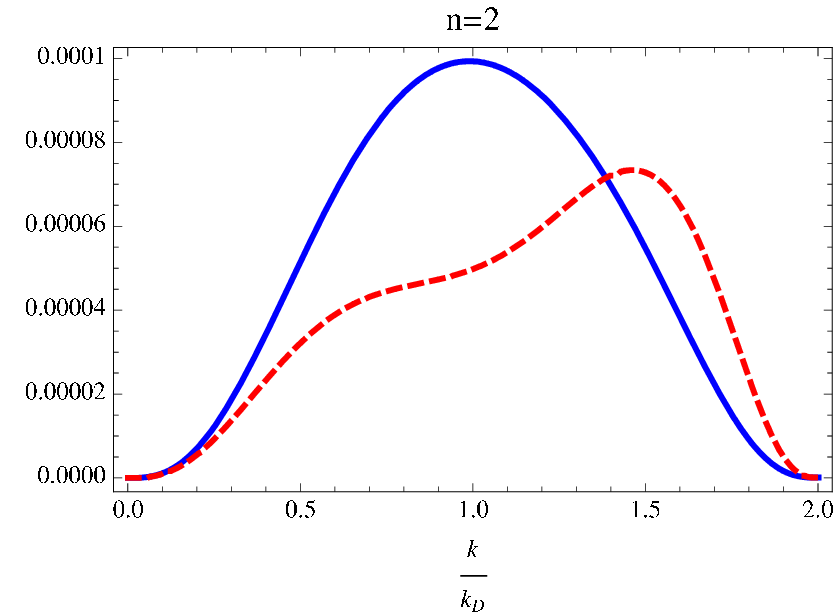}
\includegraphics[height=4cm,width=5cm]{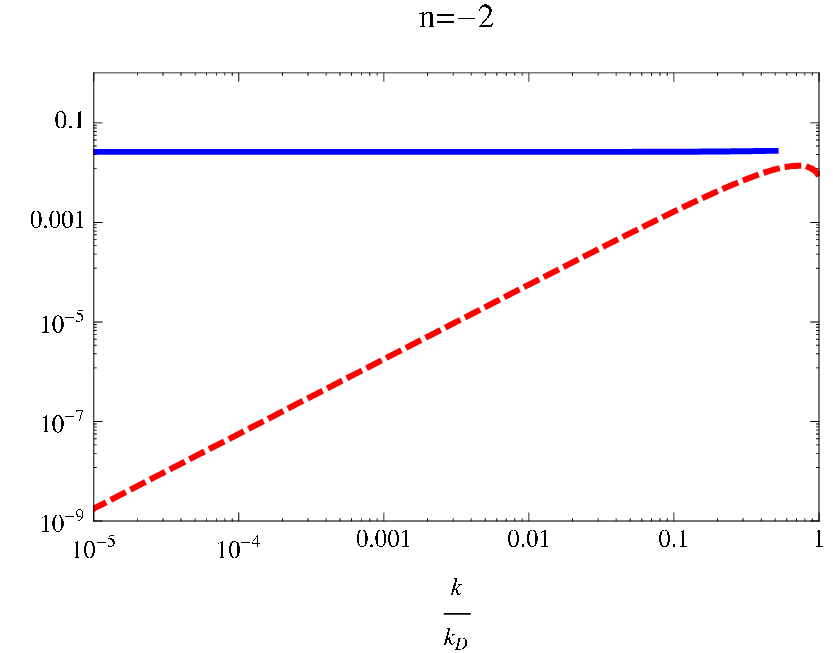}
\caption{The magnetic field bispectrum in the collinear configuration $\bp=\bq=-\bk/2$ (blue, solid) and the magnetic field spectrum to the $3/2$ (red, dashed), both multiplied by the phase space density $k^3$, as a function of $k/k_D$ for $n=2$ and $n=-2$. Note that in the $n=-2$ case, we only calculated the bispectrum up to $k=k_D/2$, while the spectrum is known up to $k=k_D$.}
\label{fig:specbispec}
\end{figure}

It is possible to calculate \eqref{collinear2} exactly for the selected values of the spectral index $n=2$ and $n=-2$. This is due to the fact that, in this configuration, the integration over the angle $\bar\phi$ becomes trivial ({\it c.f.}~ Fig.~\ref{geometry}): since $\bp=\bq$ and $\bk=-2\bq$, the integrands in \eqref{collinear2} depend only on $\cos\bar\theta$ and the boundaries given by $V_1$ and $V_2$ can be made explicit with little difficulty. The details of the calculation are given in Appendix \ref{appendixA}, while the result is shown in Fig.~\ref{figcollinear}. In the case $n=-2$ the calculation is quite involved, therefore we have evaluated only the infrared part, up to $k\leq k_D/2$. On the other hand, the case $n=2$ is simpler, and in this case we found a general, exact expression. This expression  confirms that the cutoff of the bispectrum is at $k=2k_D$, as we would expect from the analysis of the spectrum (see Fig.~\ref{fig:specbispec}), and as can be viewed easily from the last inequality of the boundary conditions of $V_1$, $|\bk+\tilde{\bk}|\leq k_D$, which shows that the maximal allowed value for $k$ is $2k_D$. 

Knowing the exact result, we can test the goodness of the approximation given in the last section at least in this configuration. Reducing the general result of \eqref{approxrhoB} in the collinear configuration, we find:
\bea
\label{approxcoll}
\left.\vev{\rho_B(\bk)\rho_B(\bq)\rho_B(\bp)}\right|_{\rm collinear}&\simeq& \frac{\de(\bk+\bp+\bq)}{144\pi^2}A^3 
\left\{ \frac{n}{2^{3n+3}(2n+3)}\left( \frac{2^{n+1}+1}{n+3}+\frac{2^{3n+4}+1}{3n+3}\right)k^{3n+3}+\frac{k_D^{3n+3}}{n+1}\right.\nonumber\\
&&\left.+\,k\rightarrow  p +k\rightarrow q
\right\}\,.
\eea
Given that $k\leq k_D$, if $n<-1$, this expression is divergent for $k\rightarrow 0$ as $k^{3n+3}$, while for $n>-1$, is it white noise. Consequently, the case $n=-2$ exhibits a divergent behaviour as $k^{-3}$, while the case $n=2$ is regular, as can be seen in Fig.~\ref{figcollinear}, where the true and approximated result are compared. In the regular $n=2$ case, the bispectrum is not pure white noise but shows a mild dependence on $k$: our approximation does not capture this dependence, but only the infrared white noise behaviour. In both cases, our approximation underestimates the true result by a factor of two. 

In Fig.~\ref{fig:specbispec}, we compare the exact result of the bispectrum in the collinear configuration with the magnetic spectrum to the power $3/2$, for $n=2$ and $n=-2$, both multiplied by the phase space density $(k/k_D)^3$. For $n=2$, they are of the same order of magnitude, as one would expect. For $n=-2$, the bispectrum goes as $k^{3n+3}$, while the spectrum as $k^{2n+3}$. The spectrum approaches the bispectrum amplitude as $k$ grows, however, the exact bispectrum has been calculated only for $k\leq k_D/2$, and this is the region shown in the plot.

\subsection{Squeezed configuration}

In the squeezed configuration one wave-vector goes 
to zero while the other two are equal but opposite in direction. 
Expliciting the case in which $\bq \simeq 0$, $\bk=-\bp$, from \eqref{rhoBsymm} we have
\bea
\left.\vev{\rho_B(\bk)\rho_B(\bq)\rho_B(\bp)}\right|_{\rm squeezed}
=\frac{\de(\bk+\bp+\bq)}{384\pi^3}& &\frac{1}{3}
\int d^3\tilde{k}\,P_{ij}(\tilde{\bk}) \left\{
P_{jl}(\bk-\tilde{\bk})[P_{il}
(\bq+\tilde{\bk})+P_{il}(\bk-\tilde{\bk})] 
\right. \nonumber \\
&&+\left. P_{jl}(\bq-\tilde{\bk})[P_{il}
(\bk-\tilde{\bk})+P_{il}(\bk+\tilde{\bk})] 
\right. \nonumber \\
&&+ \left. P_{jl}(\bk+\tilde{\bk})[P_{il}
(\bk+\tilde{\bk})+P_{il}(\bq+\tilde{\bk})] 
\right. \nonumber \\
&& \left . +\left( \bq \rightarrow \bp \simeq 0 \,, \bk \rightarrow \bq \right)
+ \left( \bq \rightarrow \bk \simeq 0 \,, \bk \rightarrow \bp \right)
\right\}\,.
\label{squeezed}
\eea
Using Eq.~\ref{productP}, and grouping the terms which are mutually equal, we find:
\bea\label{squeezed2}
& &\left.\vev{\rho_B(\bk)\rho_B(\bq)\rho_B(\bp)}\right|_{\rm squeezed}=
\frac{\de(\bk+\bp+\bq)}{384\pi^3}A^3  \times\\
& & \frac{2}{3} \left\{ \int_{V_1} d^3\tilde{k} ~ \tilde{k}^n\,\left| \bk-\tilde{\bk} \right|^n \left|\bq+\tbk\right|^n 
\left[\frac{(\hat{\tilde{k}}\cdot \bk -\tilde{k})^2}{\left| \bk-\tilde{\bk} \right|^2}+
\frac{(\hat{\tilde{k}}\cdot \bq +\tilde{k})^2}{\left|\bq+\tilde{\bk} \right|^2}+
\frac{(\bk\cdot\bq-\bk\cdot\tbk +\bq\cdot\tbk -\tilde{k}^2)[\bk\cdot\bq-(\hat{\tilde{k}}\cdot \bk)(\hat{\tilde{k}}\cdot \bq)]}{\left| \bk-\tilde{\bk} \right|^2\left|\bq+\tilde{\bk} \right|^2}\right] \right. \nonumber \\
& & + \int_{V_2} d^3\tilde{k} ~ \tilde{k}^n\,\left| \bk-\tilde{\bk} \right|^{2n} 
\left[2\,\frac{(\hat{\tilde{k}}\cdot \bk-\tilde{k})^2}{\left| \bk-\tilde{\bk} \right|^2}+
\frac{(k^2-2\,\bk\cdot\tilde{k}+\tilde{k}^2)[k^2-(\hat{\tilde{k}}\cdot \bk)^2]}{\left|\bk-\tilde{\bk} \right|^4}\right] \nonumber \\
& & + \int_{V_3} d^3\tilde{k} ~ \tilde{k}^n\,\left| \bq-\tilde{\bk} \right|^n \left|\bk-\tbk\right|^n 
\left[\frac{(\hat{\tilde{k}}\cdot \bk -\tilde{k})^2}{\left| \bk-\tilde{\bk} \right|^2}+
\frac{(\hat{\tilde{k}}\cdot \bq -\tilde{k})^2}{\left|\bq-\tilde{\bk} \right|^2}+
\frac{(\bk\cdot\bq-\bk\cdot\tbk -\bq\cdot\tbk +\tilde{k}^2)[\bk\cdot\bq-(\hat{\tilde{k}}\cdot \bk)(\hat{\tilde{k}}\cdot \bq)]}{\left| \bk-\tilde{\bk} \right|^2\left|\bq-\tilde{\bk} \right|^2}\right]  \nonumber \\
& &  \left. + \left( \bq \rightarrow \bp \simeq 0 \,, \bk \rightarrow \bq \right)
+ \left( \bq \rightarrow \bk \simeq 0 \,, \bk \rightarrow \bp \right) \right\}\,,\nonumber
\eea
where $V_1$ is given by the conditions 
\bea
\tilde{k}&\leq & k_D\nonumber \\ 
|\bk-\tilde{\bk}|&\leq& k_D \nonumber \\
|\bq+\tilde{\bk}|&\leq& k_D\,,
\eea 
$V_2$ by the conditions
\bea
\tilde{k}&\leq & k_D\nonumber \\ 
|\bk-\tilde{\bk}|&\leq& k_D \,,
\eea 
and $V_3$ by the conditions
\bea
\tilde{k}&\leq & k_D\nonumber \\ 
|\bq-\tilde{\bk}|&\leq& k_D \nonumber \\
|\bk-\tilde{\bk}|&\leq& k_D\,.
\eea 
We do not have an exact calculation of the bispectrum in the squeezed configuration. This is due to the fact that, contrary to the collinear case, the integration over the angle $\bar\phi$ is not trivial (\emph{c.f.} Fig.~\ref{geometry}). For example, for the case $\bq \simeq 0$, $\bk=-\bp$, the angle $\phi\rightarrow \pi/2$: therefore, taking for example the first integral in \eqref{squeezed2}, we see that it contains the term $(q^2+\tk^2+2\tk q \sin \bar\theta \cos\bar\phi)^{n/2}$, and the integration boundary over $\bar\theta$ and $\bar\phi$ given by $V_1$ becomes very complicated. Having already an estimation of the goodness of our approximation in the collinear case, we do not dwell on the calculation for the squeezed configuration, and use only the approximated formula. Observing the boundary conditions given by the integration volumes $V_1$, $V_2$ and $V_3$ we can however confirm that, also in the squeezed configuration, the bispectrum goes to zero at $k=q=2k_D$, since
  these are the maximally allowed values for the wave-numbers.

We reduce the general expression given in \eqref{approxrhoB} in the squeezed configuration, and find:
\bea
\label{approxsque}
\left.\vev{\rho_B(\bk)\rho_B(\bq)\rho_B(\bp)}\right|_{\rm squeezed}&\simeq& \frac{\de(\bk+\bp+\bq)}{144\pi^2}A^3 
\left\{ \frac{2n}{(n+3)(2n+3)}q^{2n+3}k^{n}+\frac{6n(n+2)}{(3n+3)(2n+3)(n+3)}k^{3n+3}\right.\nonumber\\
&&\left.+\frac{k_D^{3n+3}}{n+1}+\left( q \rightarrow p \simeq 0 \,, k \rightarrow q \right)+\left( q \rightarrow k \simeq 0 \,, k \rightarrow p \right)
\right\}\,.
\eea
For $n>-1$, the resulting white noise plateau has the same amplitude as in the collinear case. However, for $n<-1$ the divergence for $q\rightarrow 0$ is $q^{2n+3}$: therefore, it is weaker than in the collinear case, and reaches the collinear behaviour $q^{3n+3}$ only in the limit $k\rightarrow q \rightarrow 0$.

\subsection{Equilateral configuration}

In the equilateral configuration the wave-vectors form an equilateral triangle. With $\bq=k\hat{q}$, and $\bp=k\hat{p}$, using \eqref{rhoBsymm} and regrouping the equal terms one gets
\bea
\left. \vev{\rho_B(\bk)\rho_B(\bq)\rho_B(\bp)}\right|_{\rm equilateral}&=&\frac{\de(\bk+\bp+\bq)}{384\pi^3}
\, \frac{2}{3} \int d^3\tilde{k}\,P_{ij}(\tilde{\bk})\left\{ P_{jl}(\bk-\tilde{\bk})[P_{il}(k\hat{q}+\tilde{\bk})+P_{il}(k\hat{p}+\tilde{\bk})]\right.  \\
&+& P_{jl}(k\hat{q}-\tilde{\bk}) P_{il}(k\hat{p}+\tilde{\bk})\left . + \left( \bk\rightarrow \bq\,, k\hat{q}\rightarrow q\hat{k} \,, k\hat{p}\rightarrow q\hat{p} \right) + \left( \bk\rightarrow \bp\,, k\hat{q}\rightarrow p\hat{q} \,, k\hat{p}\rightarrow p\hat{k} \right)
\right\}\,. \nonumber
\label{equilateral}
\eea

Using \eqref{productP}, we can rewrite the above expression explicitly as
\bea\label{equilateral2}
& &\left.\vev{\rho_B(\bk)\rho_B(\bq)\rho_B(\bp)}\right|_{\rm equilateral}=
\frac{\de(\bk+\bp+\bq)}{384\pi^3}A^3  \times\\
& & \frac{2}{3} \left\{ \int_{V_1} d^3\tilde{k} ~ \tilde{k}^n\,\left| \bk-\tilde{\bk} \right|^n \left|k\hat{q}+\tbk\right|^n 
\left[\frac{(\hat{\tilde{k}}\cdot \bk -\tilde{k})^2}{\left| \bk-\tilde{\bk} \right|^2}+
\frac{(\hat{\tilde{k}}\cdot k\hat{q} +\tilde{k})^2}{\left|k\hat{q}+\tilde{\bk} \right|^2}+
\frac{(\bk\cdot k\hat{q}-\bk\cdot\tbk +k\hat{q}\cdot\tbk -\tilde{k}^2)[\bk\cdot k\hat{q}-(\hat{\tilde{k}}\cdot \bk)(\hat{\tilde{k}}\cdot k\hat{q})]}{\left| \bk-\tilde{\bk} \right|^2\left|k\hat{q}+\tilde{\bk} \right|^2}\right] \right. \nonumber \\
& & + \int_{V_2} d^3\tilde{k} ~ \tilde{k}^n\,\left| \bk-\tilde{\bk} \right|^n \left|k\hat{p}+\tbk\right|^n 
\left[\frac{(\hat{\tilde{k}}\cdot \bk -\tilde{k})^2}{\left| \bk-\tilde{\bk} \right|^2}+
\frac{(\hat{\tilde{k}}\cdot k\hat{p} +\tilde{k})^2}{\left|k\hat{p}+\tilde{\bk} \right|^2}+
\frac{(\bk\cdot k\hat{p}-\bk\cdot\tbk +k\hat{p}\cdot\tbk -\tilde{k}^2)[\bk\cdot k\hat{p}-(\hat{\tilde{k}}\cdot \bk)(\hat{\tilde{k}}\cdot k\hat{p})]}{\left| \bk-\tilde{\bk} \right|^2\left|k\hat{p}+\tilde{\bk} \right|^2}\right]  \nonumber \\
& & + \int_{V_3} d^3\tilde{k} ~ \tilde{k}^n\,\left| k\hat{q}-\tilde{\bk} \right|^n \left|k\hat{p}+\tbk\right|^n 
\left[\frac{(\hat{\tilde{k}}\cdot k\hat{q} -\tilde{k})^2}{\left| k\hat{q}-\tilde{\bk} \right|^2}+
\frac{(\hat{\tilde{k}}\cdot k\hat{p} +\tilde{k})^2}{\left|k\hat{p}+\tilde{\bk} \right|^2}+
\frac{(k\hat{q}\cdot k\hat{p}-k\hat{q}\cdot\tbk +k\hat{p}\cdot\tbk -\tilde{k}^2)[k\hat{q}\cdot k\hat{p}-(\hat{\tilde{k}}\cdot k\hat{q})(\hat{\tilde{k}}\cdot k\hat{p})]}{\left| k\hat{q}-\tilde{\bk} \right|^2\left|k\hat{p}+\tilde{\bk} \right|^2}\right] \nonumber \\
&& \left . + \left( \bk\rightarrow \bq\,, k\hat{q}\rightarrow q\hat{k} \,, k\hat{p}\rightarrow q\hat{p} \right) + \left( \bk\rightarrow \bp\,, k\hat{q}\rightarrow p\hat{q} \,, k\hat{p}\rightarrow p\hat{k} \right)
\right\}\,,
\eea
where again $V_1$ is given by the conditions 
\bea
\tilde{k}&\leq & k_D\nonumber \\ 
|\bk-\tilde{\bk}|&\leq& k_D \nonumber \\
|k\hat{q}+\tilde{\bk}|&\leq& k_D\,,
\eea 
and similarly for $V_2$ and $V_3$. In this case as well, we cannot solve the above integrals exactly. Like in the squeezed configuration, the integration in $d\bar\phi$ is non-trivial, since $\phi=2\pi/3$ and, for example, the first integral of \eqref{equilateral2} contains terms like $(k^2+\tk^2+2\tk k (\frac{1}{2} \sin \bar\theta \cos\bar\phi-\frac{1}{2} \cos\bar\theta))^{n/2}$. We therefore use the approximated expression in \eqref{approxrhoB}, which gives simply: 
\bea
\label{approxequi}
\left.\vev{\rho_B(\bk)\rho_B(\bq)\rho_B(\bp)}\right|_{\rm equilateral}&\simeq& \frac{\de(\bk+\bp+\bq)}{144\pi^2}A^3 
\left\{ \frac{6n}{(n+3)(3n+3)}k^{3n+3}+\frac{k_D^{3n+3}}{n+1}+\left( k \rightarrow q \right)+\left( k \rightarrow p \right)
\right\}\,.
\eea
For $n>-1$, we find again a white noise plateau of the same amplitude as in the other configurations; for $n<-1$ the divergence for $k\rightarrow 0$ is the same as in the collinear case. At first sight, this result might not seem correct: in the collinear case, in fact, by definition the wave-vectors are collinear and therefore the limits $\tilde{\bk}\rightarrow \bk$ and $\tilde{\bk}\rightarrow -\bq$ collapse into a single wave-vector configuration. In the equilateral case, on the other hand, they do not: we would therefore naively expect the same infrared behaviour of the squeezed configuration. However, the infrared divergence occurs for $k=q=p\rightarrow 0$, and in this limit $\tilde{\bk}\rightarrow \bk$ and $\tilde{\bk}\rightarrow -\bq$ are no longer distinct. Therefore, we do expect a $k^{3n+3}$ behaviour also in the equilateral case, equivalent to what we find in the collinear case and also in the squeezed one when we let not only $q$, but also $k\rightarrow 0$ (\emph{c
 .f} \eqref{approxsque}). 

We can conclude that, although it neglects the angles, the approximation in \eqref{approxrhoB} does recover the correct behaviour of the bispectrum in the analysed configurations. However, neglecting the angles certainly introduces an inaccuracy, 
because one does not account precisely for the weight with which the different configurations contribute to the total result. We were able to compare the approximated result with the exact one only in the collinear configuration, and we found an underestimation of a factor of two both for negative and positive spectral indexes. However, this does not ensure that the total, exact bispectrum is altogether only a factor of two higher than what given in \eqref{approxrhoB}, neither that it has exactly the same dependence on wave-numbers
when we significantly deviate from the  infrared limit.

\section{The CMB bispectrum}
\label{CMBbis}

Given the magnetic energy density bispectrum $\vev{\rho_B(\bk)\rho_B(\bq)\rho_B(\bp)}$, we can now evaluate the CMB bispectrum \eqref{aaa}. We use the approximated magnetic energy bispectrum \eqref{approxrhoB},
\bea
\label{I}
\vev{\rho_B(\bk)\rho_B(\bq)\rho_B(\bp)}&\simeq& \de(\bk+\bp+\bq)\frac{A^3 k_D^{3n+3}}{48\pi^2}\,\II(K,Q,P) \\
\II(K,Q,P)&=& \frac{n}{(n+3)(2n+3)}Q^n K^{2n+3} + \frac{n}{(3n+3)(2n+3)}Q^{3n+3}+\frac{n}{(n+3)(2n+3)}P^n K^{2n+3} \nonumber\\
& &+ \frac{n}{(3n+3)(2n+3)}P^{3n+3} +\frac{n}{(n+3)(2n+3)}P^n Q^{2n+3} + \frac{n}{(3n+3)(2n+3)}P^{3n+3} \nonumber\\
& &+\frac{1}{n+1}~~~~~~~~~~~~~~~~~~~~~~~~~~~{\rm for}~K\leq Q\leq P\leq 1 \,, \nonumber 
\eea
where $K=k/k_D$ and so on denote normalised wave-numbers. We want to estimate the reduced bispectrum $b_{\ell_1 \ell_2 \ell_3}$ introduced in \cite{Komatsu:2001rj}
\be
\vev{a_{\ell_1m_1}a_{\ell_2m_2}a_{\ell_3m_3}}=\mathcal{G}_{\ell_1 \ell_2 \ell_3}^{m_1 m_2 m_3}b_{\ell_1 \ell_2 \ell_3}\,,
\ee
where $\mathcal{G}_{\ell_1 \ell_2 \ell_3}^{m_1 m_2 m_3}$ is the Gaunt integral. We use the procedure described in \cite{Wang:1999vf}: starting from \eqref{aaa}, substituting in it \eqref{thetaSW} and \eqref{OmB}, and using expression (\ref{I}) for the source, we find:
\bea
b_{\ell_1 \ell_2 \ell_3}&=&\frac{\pi\,\al^3 A^3 k_D^{3n+9}}{6\,\rho_{\rm rel}^3}\int_0^\infty dx\,x^2 \int_0^1 dK\,K^2 \int_0^1 dQ\,Q^2 \int_0^1 dP\,P^2 j_{\ell_1}(Ky) j_{\ell_1}(Kx) j_{\ell_2}(Qy) \nonumber \\
& & j_{\ell_2}(Qx) j_{\ell_3}(Py)j_{\ell_3}(Px)\,\II(K,Q,P) \,, 
\eea
where $y=k_D\eta_0$ and $x=k_D r$, and $r$ comes from the decomposition of the delta function in (\ref{I}) (see \cite{Wang:1999vf}). Using the definition of the bispectrum $\II(K,Q,P)$ given in (\ref{I}), the above equation becomes
\bea
\label{bispecexplicit}
b_{\ell_1 \ell_2 \ell_3}&=&\frac{\pi\,\al^3 A^3 k_D^{3n+9}}{36\,\rho_{\rm rel}^3}  \int_0^\infty dx\,x^2 \int_0^1 dK\,K^2 j_{\ell_1}(K y) j_{\ell_1}(K x) \int_0^{K} dQ Q^2 j_{\ell_2}(Q y) j_{\ell_2}(Q x) \int_0^{Q} dP P^2 j_{\ell_3}(P y) j_{\ell_3}(P x) \nonumber\\
&\times &\left\{ a(n) \left[ K^n Q^{2n+3} +K^n P^{2n+3}+Q^n P^{2n+3} \right] + b(n) \left[2K^{3n+3}+Q^{3n+3}\right]+c(n) \right\}  \nonumber \\
&+&~{\rm permutations}\,,
\eea
where $a(n)=n/(n+3)/(2n+3)$, $b(n)=n/(3n+3)/(2n+3)$, $c(n)=1/(n+1)$, and one adds the six ordered permutations of $K$, $Q$ and $P$ which entail permutations of $\ell_1,~\ell_2,~\ell_3$.

In order to estimate the bispectrum, we substitute the upper boundaries in \eqref{bispecexplicit} with the interval $[0,1]$ in all the integrals over the momenta, since the Bessel functions peak at very low momentum: $j_{\ell_3}(Py)$ peaks at $P\simeq \ell_3/y$, and $y\gg 1$. Because of the form of the source $\II(K,Q,P)$, in (\ref{bispecexplicit}) at least one integral over the momentum is not influenced by the source. Following \cite{Fergusson:2006pr}, for each of these integral we use the approximation (\cf Eq. 6.512 of \cite{Gradstein})
\be
\int_0^1 dP\,P^2 j_{\ell_3}(P y) j_{\ell_3}(P x) \sim  \frac{1}{4}\frac{\delta(y-x)}{x^2}\,,
\ee
we then solve the integral in $dx$ using the delta function and obtain for the first term for example,
\be
\frac{a(n)}{4}\int_0^1 dK\,K^{n+2} j_{\ell_1}^2(K y)\int_0^1 dQ\,Q^{2n+5} j_{\ell_2}^2(Q y)\,,
\label{exint}
\ee
and so on. Approximate expressions for this kind of integrals are discussed in appendix~\ref{appendixAA}. 

If $n>-1$, in \eqref{bispecexplicit} we retain only the white noise term $c(n)$. All permutations give the same result in this case, and we find finally
\be
b_{\ell_1 \ell_2 \ell_3}\simeq \frac{\pi^7\,\alpha^3}{96}\frac{(n+3)^3}{n+1}\frac{\vev{B^2}^3}{\rho_{\rm rel}^3} \frac{1}{(k_D\eta_0)^4}\, ,
~~~~~~~~~~~~~~~~~~~~~{\rm for}~n>-1\,.
\ee
Values of the spectral index $n<-1$, for which the source is not pure white noise, are a bit more involved. As in the spectrum case (\cf section~\ref{sec:sachswolfe}), we cannot give a general expression valid for every $n<-1$, since the way to approximate integrals like those in (\ref{exint}) depends on the actual value of the power law exponent. Therefore, we give explicit expressions only for two values of the spectral index: $n=-2$, and $n\rightarrow -3$. Fixing the spectral index to $n=-2$, one finds
\bea
b_{\ell_1 \ell_2 \ell_3}&\simeq& \frac{\pi^8\,\alpha^3}{288}\frac{\vev{B^2}^3}{\rho_{\rm rel}^3} \frac{1}{(k_D\eta_0)^3}
\left\{\frac{1}{\ell_1}\left[\log\left(\frac{k_D\eta_0}{\sqrt{\ell_2}\sqrt{\ell_3}}\right)-\frac{2k_D\eta_0}{3\pi}\frac{1}{\ell_1}\right]+\frac{1}{\ell_2}\left[\frac{1}{2}\log\left(\frac{k_D\eta_0}{\ell_3}\right)-\frac{k_D\eta_0}{3\pi}\frac{1}{\ell_2}\right]
\right\}\nonumber\\
&+&~{\rm permutations}\, , ~~~~~~~~~~~~~~~~~~~~~~~~~{\rm for}~n=-2\,.
\eea
It is important to remark that the squeezed limit of the above expression must be taken with $\ell_3\ll \ell_2\simeq \ell_1$, since this expression has been derived from the wave-number configuration $P\leq Q\leq K$. We see that in this case, the dominant term in the bispectrum (of the order $\log(k_D\eta_0/\ell_3)$) correctly corresponds to the one coming from the dominant term in wave-number space, $P^{2n+3}$. The permutations must be treated accordingly: for example, for $Q\leq P\leq K$ one has $\ell_2\ll \ell_1\simeq \ell_3$. 

For $n\rightarrow -3$, we solve the integrals setting $n=-3$, therefore using approximation (\ref{approxmin1}) with $m=-1$, and we find then
\bea
b_{\ell_1 \ell_2 \ell_3}&\simeq &\frac{\pi^7\,\alpha^3}{288}\frac{n(n+3)^2}{2n+3}
\frac{\vev{B^2}^3}{\rho_{\rm rel}^3} \left[ \left(\frac{1}{\ell_1^2\ell_2^2}+\frac{1}{\ell_1^2\ell_3^2}+\frac{1}{\ell_2^2\ell_3^2}\right)+\frac{\pi}{16} \frac{n+3}{n+1}\,k_D\eta_0\left( \frac{1}{\ell_1^5}+\frac{1}{2\ell_2^5}\right)\right]\nonumber\\
&+&~{\rm permutations}\, ,~~~~~~~~~~~~~~~~~~~~~~~~~{\rm for}~n\approx -3\,,
\label{bispecnm3}
\eea
where the same considerations as above apply for the squeezed limit. The second term in the above expression, coming from the term proportional to $b(n)$ in \eqref{bispecexplicit} is sub-leading, since it contains a factor $n+3$. Note that since $\vev{B^2}\propto (n+3)^{-1}$, the leading term of the bispectrum diverges in $n \rightarrow -3$ as $(n+3)^{-1}$, like the spectrum (\cf \eqref{specnmen3mezzi}): this divergence is connected to the infrared divergence of the magnetic energy \footnote{The above expression is valid only for $n\rightarrow -3$ so the denominator is always finite. Note however that the apparent divergence for $n=-3/2$ is just an artefact due to our approximation (\cf \eqref{rhoBmen3mezzi}): $n=-3/2$ would correspond to a threshold value for which $|\rho_B (k)|^2$ diverges logarithmically for $k \rightarrow 0$ and is not simply white noise.}. 

The leading term of the above result \eqref{bispecnm3}, reduced to the squeezed and the equilateral configurations, gives the same result as found in \cite{Seshadri:2009sy} (\cf eqs.~(17) and (18) and discussion thereafter, we remind that we use $\al=0.1$).

\section{Estimation of the signal}
Since the signal-to-noise ratios $(S/N)$ we will be interested in is  some function 
of the maximum multipole a given 
experiment can reach, $\ell_{\rm max}\gg 1$, we can use the flat-sky approximation ~\cite{BZ,Hulensing} and write
for the bispectrum 

\be
\langle a(\vec{\ell}_1)a(\vec{\ell}_2)a(\vec{\ell}_3) \rangle 
   = (2\pi)^2\delta^{(2)}(\vec{\ell}_{123}) B(\ell_1,\ell_2,\ell_3)\, ,
\ee 
where $\vec{\ell}_{123}=\vec{\ell}_1+\vec{\ell}_2+\vec{\ell}_3$. With this notation,  
the reduced bispectrum $b_{\ell_1\ell_2\ell_3}$
coincides with the
bispectrum $B(\ell_1,\ell_2,\ell_3)$. 

Our goal now is to quantify the level of NG coming from the stochastic magnetic field  
and eventually to give a bound on the
amplitude of the magnetic field. One way to do it is to exploit the present bound on the primordial
local non-Gaussianity parametrized by the quantity $f^{\rm loc}_{\rm NL}$. As we mentioned in the introduction, the
search for a non-vanishing bispectrum of a local type  has given so far a null result and  currently  $f^{\rm loc}_{\rm NL}$
is bounded in the range $-9<f^{\rm loc}_{\rm NL}<111$. As the shape of the non-Gaussian signature from the
stochastic magnetic field may be  different from the one of the local type, one may not  directly apply the
bounds coming from WMAP5 whose search for non-Gaussianity is optimised to search for
local primordial contribution. Instead, we proceed in the following way. First, we  
define the Fisher matrix (see, for example, \cite{Komatsu:2001rj})
\begin{equation} 
F_{ij}=\frac{f_{\rm sky}}{(2\pi)^2\pi}\int d^2 \ell_1 d^2 \ell_2  d^2 \ell_3 
\,\delta^{(2)}(\vec{\ell}_{123})\,\frac{
B_{i}(\ell_1,\ell_2,\ell_3)\, B_{j}(\ell_1,\ell_2,\ell_3)}{6\, C(\ell_1)\,C(\ell_2)\, C(\ell_3)}\, ,
\end{equation}
where $f_{\rm sky}$ is the portion of the observed-sky in a given experiment and 
$i$ (or $j$)$=({\rm mag},{\rm loc})$. The first entry $F_{\rm mag,mag}$ 
of the Fisher matrix corresponds to 
the signal-to-noise ratio $(S/N)^2$ 
provided by the stochastic magnetic field to the non-Gaussianity. We have defined the power spectrum
in  the flat-sky approximation by $\langle a(\vec{l}_1)a(\vec{l}_2) \rangle 
   = (2\pi)^2\delta^{(2)}(\vec{l}_{12}) C(\ell_1)$ with $\ell^2\,C(\ell)={\cal A}/\pi$ and ${\cal A}\simeq 17.46 \times 10^{-9}$ 
is the amplitude of the  primordial gravitational potential 
power spectrum computed at first-order. In other words, we assume that the two-point correlation function
is dominated by the usual adiabatic contribution from inflation. 
Finally, the local bispectrum is given by \cite{BZ}
\begin{equation}
\label{eq:loc}
B_{\rm loc}(\ell_1,\ell_2,\ell_3) = \frac{2\,f_{\rm NL}^{\rm loc} \,  
 {\cal A}^2}{\pi^2} \left(\frac{1}{\ell_1^2\ell_2^2}+\,{\rm cycl.}  \right) \, .
\end{equation}
Notice that all these expressions are obtained in the Sachs-Wolfe approximation. We will return back to this
point shortly.

Next, we define an effective   $f^{\rm eff}_{\rm NL}$ which minimises the $\chi^2$ defined as 
\begin{equation}
\chi^2=\int d^2 \ell_1 d^2 \ell_2  d^2 \ell_3 
\,\delta^{(2)}(\vec{\ell}_{123})\,
\frac{\left(\left. f^{\rm eff}_{\rm NL}\,B_{\rm loc}(\ell_1,\ell_2,\ell_3)\right|_{f_{\rm NL}^{\rm loc}=1}-
B_{\rm mag}(\ell_1,\ell_2,\ell_3)\right)^2}{6\, C(\ell_1)\,C(\ell_2)\, C(\ell_3)}\, . \nonumber 
\end{equation}
One finds 
\begin{equation}
\label{effr}
f^{\rm eff}_{\rm NL}= \frac{F_{\rm mag,loc}}{F_{\rm loc ,loc}} \Big|_{f_{\rm NL}^{\rm loc}=1}\,.
\ee
The signal-to-noise ratio for the primordial local case  has already been computed in the flat-sky approximation in 
Ref.~\cite{BZ}. The result is that $F_{\rm loc ,loc}\simeq(4/\pi^{2})f_{\rm sky}{\cal A} (f_{\rm NL}^{\rm loc})^2\,
\ell_{\rm max}^2\,
\log(\ell_{\rm max}/\ell_{\rm min})$. The logarithm is typical of scale invariant power spectra and  $\ell_{\rm min}$ is the 
minimum multipole compatible with the flat-sky approximation. The physical meaning of $f^{\rm eff}_{\rm NL}$ is the following:
it is the best 
value of  the local $f^{\rm loc}_{\rm NL}$ 
which best mimics the bispectrum from a stochastic magnetic field background. As such, we can
apply to this value  the current observational limits. 

We start with the simplest case $n\approx -3$. Indeed, for $n$ close to $-3$, the leading term of the bispectrum is of the same form of the local primordial bispectrum (\ref{eq:loc}) in the squeezed limit $\ell_3\ll \ell_1\simeq \ell_2$ and we immediately find

\be
f^{\rm eff}_{\rm NL}\simeq \frac{3\pi^9\,\alpha^3}{288\,{\cal A}^2}\frac{n(n+3)^2}{2n+3}
\frac{\vev{B^2}^3}{\rho_{\rm rel}^3}\simeq 10^{-2}\,(n+3)^2\,
\left(\frac{\vev{B^2}}{(10^{-9}\Gauss)^2}\right)^3\,
,~~~~~~~~~~~~{\rm for}~n\approx -3\, .
\ee 
In the case in which the bispectrum is independent from the multipoles, that is for  $n>-1$, we find 

\be
f^{\rm eff}_{\rm NL}\simeq \frac{\pi^9\,\alpha^3}{2304\,{\cal A}^2}\frac{(n+3)^3}{n+1}\frac{\vev{B^2}^3}{\rho_{\rm rel}^3} 
\left(\frac{\ell_{\rm max}}{\ell_D}\right)^4\,
\frac{1}{\log(\ell_{\rm max}/\ell_{\rm min})}\simeq 
6\times 10^{-7}\,\frac{(n+3)^3}{n+1}\,
\left(\frac{\vev{B^2}}{(10^{-9}\Gauss)^2}\right)^3\,
,~~~~~~~{\rm for}~n> -1\, .
\ee
Finally, for the case $n=-2$, we find

\be
f^{\rm eff}_{\rm NL}\simeq \frac{5\pi^{10}\,\alpha^3}{2304 \,{\cal A}^2}\frac{\vev{B^2}^3}{\rho_{\rm rel}^3} 
\left(\frac{\ell_{\rm max}}{\ell_D}\right)^3\,
\frac{\log(\ell_{D}/\ell_{\rm max})}{\log(\ell_{\rm max}/\ell_{\rm min})}
\simeq 
 5\times 10^{-5}\,
\left(\frac{\vev{B^2}}{(10^{-9}\Gauss)^2}\right)^3\,
,~~~~~~~~~~~~{\rm for}~n= -2\, .
\ee
In all numerical estimates we have taken 
$\ell_D=k_D\eta_0\simeq 3000$, $\ell_{\rm max}\sim 750$, $\ell_{\min}\sim 10$, $\al\simeq 0.1$, and \eqref{Omtot}. We
see that the effective value of non-Gaussianity  $f^{\rm eff}_{\rm NL}$ is  smaller 
than the  present upper bound of ${\cal O}(10^2)$ on $f^{\rm loc}_{\rm NL}$ \cite{wmap5} for magnetic fields
${\cal O}(10)\cdot 10^{-9}$ Gauss for $n\approx -3$ and ${\cal O}(20)\cdot 10^{-9}$ Gauss for the other cases\footnote{ 
We have obtained similar estimates repeating the same procedure to define an effective
non-Gaussianity parameter starting from a primordial equilateral configuration for which WMAP5 limits exist. In such a case
the primordial equilateral configuration is peaked for $\ell_1\sim\ell_2\sim\ell_3$ and the effective
non-Gaussianity parameter scales with $\ell_{\rm max}$ with one power less than the corresponding one obtained
from a local primordial bispectrum.}. 

Accounting more precisely for the value of the damping scale $k_D$ as a function of the spectral index and of the magnetic field amplitude using \eqref{kD}, we obtain 
\bea
\sqrt{\vev{B^2}}\leq 9~{\rm nGauss~} & {\rm for}~n=-2.9 \nonumber\\
\sqrt{\vev{B^2}}\leq 25~{\rm nGauss} & {\rm for}~n=-2 \nonumber\\ 
\sqrt{\vev{B^2}}\leq 20~{\rm nGauss} & {\rm for}~n=2
\eea
The corresponding bound on the magnetic field  amplitude $B_\la$ (\cf \eqref{Bla}) on the scale $\la=0.1$ Mpc is unchanged for $n\rightarrow -3$, it becomes $B_\la\leq 26$ nGauss for $n=-2$, and is less stringent as $n$ grows, becoming irrelevant for $n=2$: $B_\la\leq 2$ $\mu$Gauss. This is a consequence of the fact that the procedure of using an effective $f_{\rm NL}$ returns a bound on the integrated magnetic field spectrum, and therefore for very blue spectra the constraint on large scales is irrelevant.

A word of caution  is in order here though. In all our estimates, 
 we have used the Sachs-Wolfe approximation for all bispectra. This is certainly a sufficiently good approximation
for an experiment like WMAP whose maximum multipole is $\ell_{\rm max}\sim 750$. 
This is because  the transfer functions for both the scalar contribution to the CMB anisotropies from the stochastic magnetic
field and the one from the inflationary adiabatic modes may be taken roughly equal to unity up to $\ell\sim 750$ and they
do not affect the computation of the Fisher matrix elements, see \cite{BZ,br}. 
However, for higher multipoles, say $\ell\sim 2000$, typical of an experiment like Planck, the inclusion of the
transfer functions will be crucial because the anisotropies from the adiabatic inflationary modes get an exponential
suppression due to the Silk damping, while the ones from the scalar modes from the stochastic magnetic
field show a much milder suppression \cite{Finelli:2008xh,PFP}. This will increase the value of  $f^{\rm eff}_{\rm NL}$. Needless to say,
the inclusion of the vector and tensor contributions from the magnetic field will help to increase the non-Gaussian signal too.

While writing this paper, the preprint \cite{Seshadri:2009sy} appeared where the computation of the bispectrum from a stochastic
magnetic field background was presented for  the case $n\approx -3$. Our findings agree with those
in Ref. \cite{Seshadri:2009sy} and extend them to other values of the spectral index and by 
the estimation of the signal-to-noise ratio and of the effective non-Gaussianity
parameter. 

\acknowledgments

CC wishes to thank Iain Brown, Ruth Durrer, Martin Kunz, Roy Maartens, Antti V\"{a}ihk\"{o}nen and Filippo Vernizzi for helpful discussions. CC acknowledge support from the funding INFN IS PD51 for visiting IASF Bologna. 

\appendix

\section{Integrals of Bessel functions}
\label{appendixAA}

In order to evaluate both the magnetic field spectrum and bispectrum at large angular scales, we need to evaluate integrals of the type
\be
\label{exintegral}
\int_0^y dx\,x^{m} j_{\ell}^2(x)
\ee
with $y\gg 1$. This integral can be expressed generically in terms of hypergeometric functions; however, good approximations can be found, which are much simpler. 

For $m=2$, the integral can be performed exactly:  one has
\be
\label{intspectrum}
\int_0^{y} dx\,x^2\,j^2_\ell(x)=\frac{\pi}{4}y^2\left[ J^2_{\ell+\frac{1}{2}}(y)-\frac{2}{y}\big(\ell+\frac{1}{2}\big)
J_{\ell+\frac{1}{2}}(y) J_{\ell+\frac{3}{2}}(y)+J^2_{\ell+\frac{3}{2}}(y)\right]\simeq \frac{y}{2}
\ee
where since $y\gg \ell$ we used the expansion of the Bessel functions for large arguments. 

For $m<1$, the integral reaches a constant value for $y\gg \ell$, and can therefore be evaluated in the limit $y\rightarrow \infty$. We find 
\bea
\label{approxmin1}
\int_0^y dx\,x^{m} j_{\ell}^2(x)&\simeq&\frac{1}{4}\left[\frac{\sqrt{\pi}\, \Gamma(\frac{1-m}{2})\Gamma(\ell+\frac{m+1}{2})}{\Gamma(1-\frac{m}{2})\Gamma(\ell+\frac{3-m}{2})}+y^{m-2}\left(\frac{2y}{m-1}+\sin(\pi\ell-2y)\right)\right]~~\stackrel{\ell\gg 1}{\longrightarrow}~~ \frac{\sqrt{\pi}\,\Gamma(\frac{1-m}{2})}{4\,\Gamma(1-\frac{m}{2})}\,\ell^{m-1} \nonumber\\
&&{\rm for}~m<1\,,~y\gg \ell
\eea

The case $m=1$ is a bit more involved: the integral (\ref{exintegral}) grows logarithmically with $y$ and cannot be evaluated with the same approximation as before. In this case we set
\be
\label{approx1}
\int_0^y dx\,x\, j_{\ell}^2(x) \simeq \int_\ell^y \frac{dx}{x}\,\cos^2\left(x-\frac{\pi}{2}\ell-\frac{\pi}{4}\right)\simeq \frac{1}{2}\left[\log(y)-\log(\ell)\right]~~~~~~~~~{\rm for}~y\gg \ell\,.
\ee
We are neglecting the subdominant contribution to the integral of the interval $[0,\ell]$, therefore this approximation is slightly underestimating the true result. However, it captures the correct behaviour in $\ell$ and $y$. These  approximations are shown in Fig.~\ref{fig:approxbessel}.

\begin{figure}
\includegraphics[height=4cm,width=5cm]{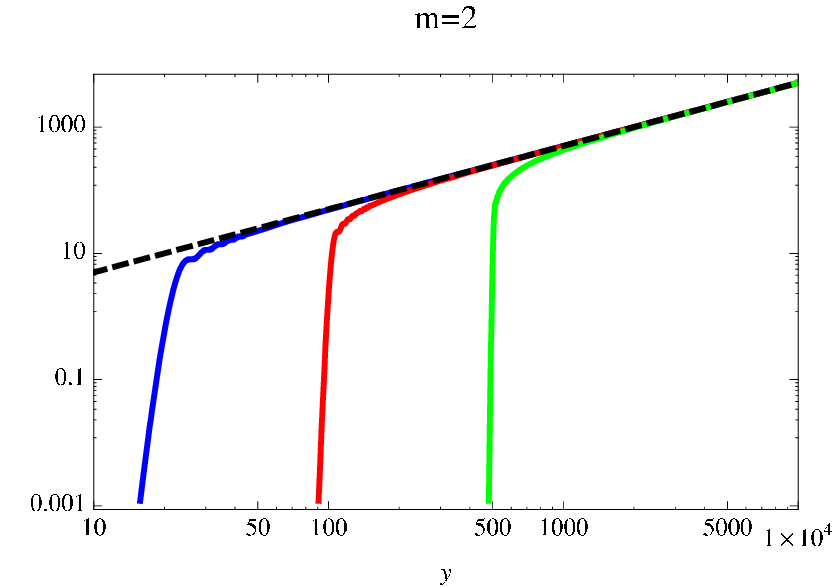}
\includegraphics[height=4cm,width=5cm]{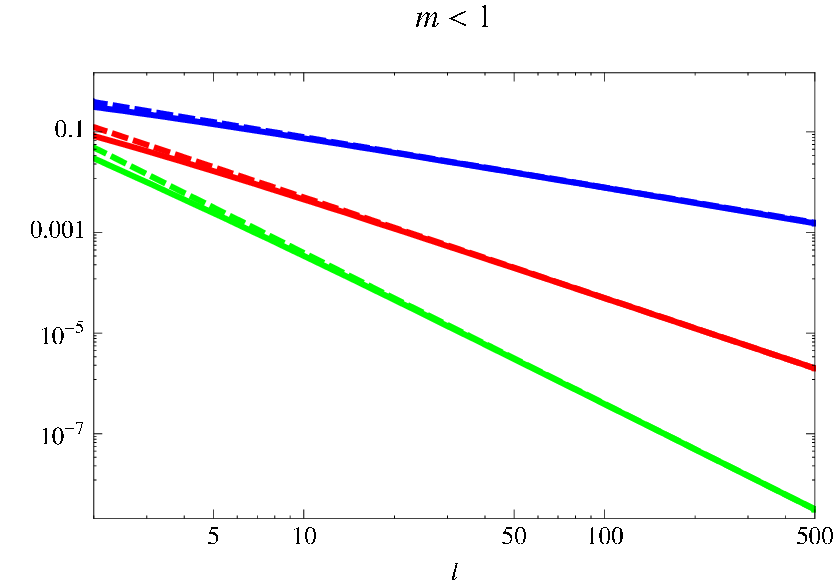}\\
\includegraphics[height=4cm,width=5cm]{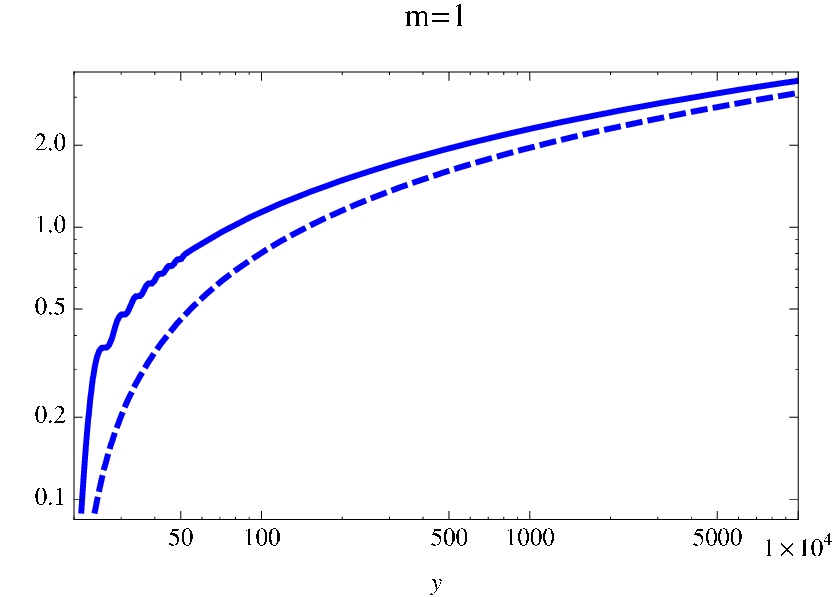}
\includegraphics[height=4cm,width=5cm]{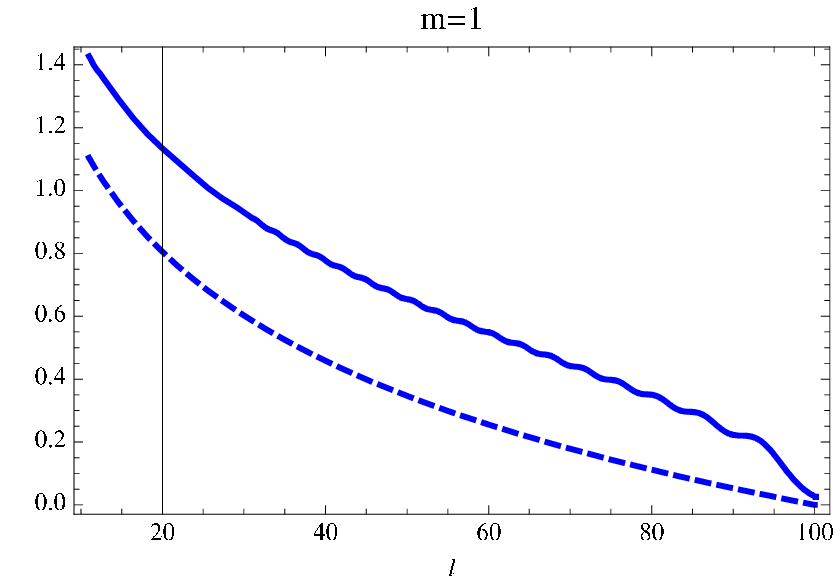}
\caption{The approximations for the integral in \eqref{exintegral}. Upper left plot, for $m=2$: the integral (solid) and the approximation $y/2$ (dashed) are shown for $\ell=20$, $\ell=100$, $\ell=500$ as a function of $y$. Upper right plot, for $m<1$: the approximations for $y\gg\ell$ (solid) and for $\ell\gg 1$ (dashed) given in \eqref{approxmin1} are shown as a function of $\ell$ for $m=0$, $m=-1$ and $m=-2$. Lower plots, for $m=1$: the integral (solid) and the approximation in \eqref{approx1} (dashed) are shown as a function of $y$ for $\ell=20$ (left plot) and as a function of $\ell$ for $y=100$ (right plot).}
\label{fig:approxbessel}
\end{figure}

\section{Bispectrum in Collinear Configuration}
\label{appendixA}

In the following appendix we describe the technique used to calculate the
magnetic energy density bispectrum in the collinear configuration
\eqref{collinear}.

Due to the complexity of the calculations we restrict to
analytical solutions of the bispectrum integral for two representative spectral
indexes: the case $n=2$ (the typical
spectrum of a magnetic field generated by a causal mechanism), and the case $n=-2$ (in order to
investigate the behaviour of the spectrum also for negative spectral
indexes).

The magnetic energy density bispectrum in the collinear configuration is
given in \eqref{collinear}. From this we extract the integral in the
momenta which is given by the permutation over the three momenta
$\bf{K},\bf{P},\bf{Q}$ of three basic integrals\footnote{For simplicity of notation in this appendix we use re-scaled
 variables: $K=k/k_D$, $Q=q/k_D$, $P=p/k_D$ and $\tK=\tk/k_D$}:
\be
I(K)=\int d\tK \int dx (I_a(K,\tK)+I_b(K,\tK)+I_c(K,\tK))\,,
\ee
where $x=\hat{K}\cdot\hat{\tilde{K}}$.
The functions $I_a(K,\tK),I_b(K,\tK),I_c(K,\tK)$ are:
\ba
I_a(K,\tK)&=& \tK^{2 + n} (\frac{K^2}{4} + \tK^2 + K \tK x)^{\frac{n}{2}} (K^2+\tK^2 + 2 K \tK x)^{-1 + \frac{n}{2}}\nonumber\\
&& \Big(\frac{8\tK^4 + 24K\tK^3 x + K^4 (1 + x^2) + 3 K^3 \tK x (3 + x^2) +  K^2 \tK^2 (7 + 19 x^2)}{K^2 + 4 \tK^2 + 4 K \tK x}\Big)\\
I_b(K,\tK)&=& \tK^{2 + n} (\frac{K^2}{4} + \tK^2 + K \tK x)^{n/2} (\tK^2 +\frac{1}{4} K (K - 4 \tK x))^{n/2}\nonumber\\
&& \Big(\frac{(32 \tK^4 + 4 K^2 \tK^2 (1 - 5 x^2) + K^4 (1 + x^2))}{((K^2 +4 \tK^2)^2 - 16 K^2 \tK^2 x^2)}\Big)\\
I_c(K,\tK)&=& \tK^{2 + n} (K^2 + \tK^2 - 2 K \tK x)^{-1 +\frac{n}{2}} (\tK^2 +\frac{1}{4} K (K - 4\tK x))^{n/2}\nonumber\\
&&\Big( \frac{(8 \tK^4 - 24 K \tK^3 x + K^4 (1 + x^2) - 3 K^3 \tK x (3 + x^2) + K^2 \tK^2 (7 + 19 x^2))}{(K^2 + 4 \tK^2 - 4 K \tK x)}\Big)
\ea
We note that due to the symmetry $\tK\rightarrow -\tK$ we have that the first
and the third integrals are indeed the same $I_a(K)=I_c(K)$, therefore to
obtain the energy density bispectrum in the collinear configuration we need to
solve only the two integrals of $I_a(K,\tK)$ and $I_b(K,\tK)$.
\newline

\subsection{Integration Domains}
The sharp cut-off of the PMF spectrum at the damping scale $k_D$, imposed to account for the magnetic
fields suppression on small scales, leads to many
conditions on the angle $\hat{\tK}\cdot \hat{K}$. This causes the integration
domain to be  split into various sub-domains.
The conditions are different for $I_a$ and $I_b$, therefore for simplicity in the following we consider the two integrations separately.
\subsection{Domains of $I_a$}
The sharp cut off imposes:
\ba
\tK<1\nonumber\\
(\frac{K^2}{4} + \tK^2 + K \tK x)<1\nonumber\\
(K^2+\tK^2 + 2 K \tK x)<1\nonumber
\ea
This leads to the following integration scheme:
\begin{eqnarray}
1)&&  0<K<1 \nonumber\\
&&\int_{0}^{1-K}d\tK \int_{-1}^{1}dx\,I_a(\tK,K) 
+ \int_{1-K}^{1}d\tK\int_{-1}^{\frac{1 - K^2 - \tK^2}{2 K \tK}}dx\, I_a(\tK,K)
\nonumber\\
2) && 1<K<2 \nonumber\\
&& \int_{K-1}^{1}d\tK 
\int_{-1}^{\frac{1 - K^2 - \tK^2}{2 K \tK}} dx\, I_a(\tK,K)
\label{intscheme}
\end{eqnarray}

\subsection{Domains of $I_b$}
The sharp cut off imposes:
\ba
\tK<1\nonumber\\
(\frac{K^2}{4} + \tK^2 + K \tK x)<1\nonumber\\
(\frac{K^2}{4} + \tK^2 - K \tK x)<1\nonumber
\ea
This leads to the following integration scheme for $0<K<2$:
\begin{eqnarray}
\int_{0}^{\frac{2-K}{2}}d\tK \int_{-1}^{1}dx\,I_b(\tK,K) 
+ \int_{\frac{2-K}{2}}^{\frac{\sqrt{4-K^2}}{2}}d\tK\int_{\frac{-1 + K^2/4 + \tK^2}{K \tK}}^{\frac{1 - K^2/4 - \tK^2}{K \tK}}dx\,I_b(\tK,K)
\nonumber\\
\end{eqnarray}
in the interval $\frac{\sqrt{4-K^2}}{2}<\tK<1$ the integral collapses to zero.

\subsection{n=2}
First we consider the case $n=2$ which is the easiest from the
point of view of the calculations.
In fact the angular integrand functions for this spectral index simply reduce to:
\begin{eqnarray}
I_a(K,\tK,x)&=&\frac{1}{4} \tK^4 (8 \tK^4 + 24 K \tK^3 x + K^4 (1 + x^2) + 3 K^3
\tK x (3 + x^2) + K^2 \tK^2 (7 + 19 x^2))\nonumber\\
I_b(K,\tK,x)&=&\frac{1}{16} \tK^4 (32 \tK^4 + 4 K^2 \tK^2 (1 - 5 x^2) + K^4 (1 + x^2))
\end{eqnarray}
Once performed the angular integrations, following the integration  scheme reported in the
previous paragraph, the radial integrations become trivial and the result is:
\be
I(K)|_{n=2}=\Big(\frac{4}{3} - 3 K + \frac{20 K^2}{7} - \frac{23 K^3}{16} +\frac{2 K^4}{5} -\frac{K^5}{16} +\frac{K^7}{256} -\frac{17 K^9}{53760}\Big)
\ee
In Fig.~\ref{figcollinear} we have shown the result for $n=2$.
We note that, as it happens for the energy density spectrum, also the
PMF energy density bispectrum goes to zero for $K=2$ as expected.

\subsection{n=-2}
Here we consider the case $n=-2$.
The functions $I_a$ and $I_b$ for this spectral index reduce to:
 \begin{eqnarray}
I_a(K,\tK,x)&=&\frac{4 (8 \tK^4 + 24 K \tK^3 x + K^4 (1 + x^2) + 3 K^3 \tK x (3 + x^2) + 
   K^2 \tK^2 (7 + 19 x^2))}{(K^2 + \tK^2 + 2 K \tK x)^2 (K^2 + 4 \tK^2 + 4 K \tK x)^2}\nonumber\\
I_b(K,\tK,x)&=&\frac{16 (32 \tK^4 + 4 K^2 \tK^2 (1 - 5 x^2) + K^4 (1 + x^2))}{((K^2 + 4 \tK^2)^2 - 16 K^2 \tK^2 x^2)^2}
\end{eqnarray}
We note how these functions are far more complicated than the ones for the
$n=2$ case.
Once performed the angular integrations in both the integrals we have the
appearance of absolute values like $|K-2\tK|$ and $|K-\tK|$, their presence
influences the integration domains creating further splitting into several
sub-domains.
 Since we are interested in the effect on CMB where only the low $K$ part of the spectrum has a role we
 restrict ourselves to the $K<1/2$ region of the spectrum.
The analytical result for $n=-2$ unfortunately has a very long and complicated
form, therefore, for the sake of simplicity, we show only the infrared
limit:
\begin{eqnarray}
I_a(K)\sim \frac{24.674}{K^3}\nonumber\\
I_b(K)\sim \frac{24.674}{K^3}\nonumber\\
I(K)\sim\frac{73.8367}{K^3}\nonumber
\end{eqnarray}
Fig.~\ref{figcollinear} shows the exact result.

\end{document}